\documentclass{article}
\usepackage{spconf,amsmath,graphicx}
\usepackage{amsfonts}

\usepackage{multirow}
\usepackage{multicol}
\usepackage{mathtools}

\usepackage{amsmath,epsfig}
\usepackage{amsfonts}
\usepackage{url}
\usepackage{amssymb}
\usepackage{amsthm}

\usepackage{caption}
\usepackage{subcaption}

\usepackage{algorithm}
\usepackage{algpseudocode}

\def\x{{\mathbf x}}
\def\y{{\mathbf y}}
\def\n{{\mathbf n}}
\def\z{{\mathbf z}}
\def\u{{\mathbf u}}
\def\v{{\mathbf v}}
\def\d{{\mathbf d}}
\def\A{{\mathbf A}}
\def\H{{\mathbf H}}
\def\J{{\mathbf J}}
\def\U{{\mathbf U}}
\def\V{{\mathbf V}}
\def\C{{\mathbf C}}
\def\R{{\mathbb R}}

\title{Exploring the solution space of linear inverse problems with GAN latent geometry}
\name{Antonio Montanaro, Diego Valsesia, Enrico Magli}
\address{Department of Electronics and Telecommunications -- Politecnico di Torino, Italy}
\begin{document}
\ninept
\maketitle
\begin{abstract}
Inverse problems consist in reconstructing signals from incomplete sets of measurements and their performance is highly dependent on the quality of the prior knowledge encoded via regularization. While traditional approaches focus on obtaining a unique solution, an emerging trend considers exploring multiple feasibile solutions. In this paper, we propose a method to generate multiple reconstructions that fit both the measurements and a data-driven prior learned by a generative adversarial network. In particular, we show that, starting from an initial solution, it is possible to find directions in the latent space of the generative model that are null to the forward operator, and thus keep consistency with the measurements, while inducing significant perceptual change. Our exploration approach allows to generate multiple solutions to the inverse problem an order of magnitude faster than existing approaches; we show results on image super-resolution and inpainting problems.
\end{abstract}

\vspace*{-5pt}
\section{Introduction}
\label{sec:intro}
\vspace*{-5pt}
Linear inverse problems are ubiquitous in the sciences as they are tasked with reconstructing a signal of interest from a set of typically incomplete or degraded measurements. In the imaging field alone \cite{bertero2021introduction}, numerous problems of interest such as deblurring, super-resolution, inpainting, compressed sensing, and many more fit this framework. Due to the ill-posed nature of the problem, one needs strong regularization to find reconstructions that fit the measurements and the a priori knowledge of the signal properties. Traditional approaches focused on hand-crafting regularizers to yield a unique solution by casting reconstruction as a convex optimization problem \cite{mairal2009online}. However, one must accept that the quality of this unique solution can only be as good as how well the chosen regularizer function captures the signal properties. For this reason, recently, data-driven methods based on neural networks \cite{lucas2018using,monga2021algorithm} started learning priors directly from the complex distributions of the signals of interest, resulting in improved reconstruction capabilities.

Nevertheless, even when using data-driven priors, we can hardly hope to capture a perfect model of our signals of interest, which in turn affects how faithful the reconstructed signal is to the true one that generated the measurements. For this reason, a novel paradigm is emerging where multiple feasible reconstructions are generated, in an effort to boost interpretability of the inversion process and expose the biases of the models.

In this paper, we use generative adversarial networks (GANs) as priors modeling the distribution of our signals of interest. We present a geometrical perspective on the latent space of such models, which allows to explore the solution space of a linear inverse problem. By exploration of the solution space, we mean finding \textit{multiple reconstructions} that are consistent with the measurements but also consistent with the model of the data distribution. We show that it is possible to modify an initial solution by moving towards directions in the latent space that are ``null'' with respect to the measurements operator (i.e., they do not significantly perturb the measurements) while inducing semantic change. Our proposed technique, called e-GLASS (exploring GAN LAtent Space Solutions), is general as it can be applied to any linear inverse problem and is an order of magnitude faster than state-of-the-art methods such as PULSE \cite{menon2020pulse} which generate multiple solutions by solving an optimization problem from different random initializations.

\vspace*{-5pt}
\section{Background and related work}
\label{sec:bkg}
\vspace*{-5pt}
Let us start from a general linear forward model of the form: 
\vspace*{-5pt}
\begin{align}\label{eq:model}
\y = \A\x + \n
\end{align}
where $\y \in \R^{m}$ is a noisy observation from an unknown signal $\x \in \R^{n}$ with $m \leq n$, depending on the specific problem; $\n$ is an additive noise and $\A$ is a measurement matrix.

As an example, $\y$ can be a degraded image, e.g., with low resolution or blurred, and we want to reconstruct the image $\x$ starting from the measurements $\y$. However, this problem is ill-posed as there can be infinitely many solutions satisfying the measurements, or even none due to noise. A large body of work has been devoted to the development of priors to model $\x$ as accurately as possible to regularize the problem towards admitting a unique solution. Such works frame reconstruction as a Maximum a Posteriori (MAP) estimation problem, with the unique solution obtained by solving
\begin{align*}
    \hat{\x} = \arg\min_{\x} \Vert \y - \A\x \Vert_2^2 + \lambda R(\x),
\end{align*}
for some handcrafted regularizer $R$ encoding the prior. This line of works typically defines priors such that the reconstruction problem is convex with a unique global minimum. This means that a single solution to the problem can be obtained, whose properties are strictly intertwined with the ability to craft a suitable prior $R$.

New recent approaches involve generative models, such as GANs, to learn priors in a data-driven fashion \cite{creswell2018inverting}, \cite{xia2021gan}. A GAN learns a function $G$ that maps a latent vector $\z$ into a sample from the data distribution. The popular approach of GAN inversion solves inverse problems by seeking the latent vector $\hat{\z}$ that best fits the measurements $\y$. This is done by minimizing the distance between $\y$ and the degraded version of the generated data $G(\hat{\z})$, under the forward model $\A$: 
\begin{align}\label{eq:stageI_loss}
\hat{\z} = \arg\min_{\z} \Vert \y - \A G(\z) \Vert_{2}^{2},\quad \hat{\x} = G(\hat{\z}).
\end{align}
Unlike convex optimization methods with handcrafted priors, GAN inversion is non-convex due to the use of neural networks, thus admitting multiple local minima. 

While most works have used GAN inversion to generate a single solution to the inverse problem, there has been recent growing interest  in exploring the \textit{solution space} of inverse problems, i.e., finding multiple solutions, among the infinitely many possible, that are consistent with the measurements and some data prior. The seminal work on this topic is PULSE \cite{menon2020pulse}, which uses GAN inversion to super-resolve low-resolution faces. PULSE generates multiple plausible solutions by solving Eq. \eqref{eq:stageI_loss} via gradient descent, and starting from different random guesses of $\z$. Due to the non-convex nature of the optimization, different solutions \textit{may} be reached when starting from different initializations. The main drawback of PULSE lies in its complexity, requiring to solve an optimization problem for each solution and the lack of any guarantee that a different initialization will converge to a different minimum.
Other works \cite{lugmayr2021ntire} have sought to generate multiple solutions for the super-resolution problem, but they lack generality and can only be applied to very specific neural networks devised only for the super-resolution task.

Finally, we remark that there is extensive literature on GAN editability \cite{9241434}, \cite{pan2021exploiting}, \cite{hussein2020image}, \cite{shen2020interpreting}, seeking to manipulate the latent space of GANs to induce semantically interesting transformations. However, such works are not in the framework of solutions to inverse problems and are not concerned with fidelity with measurements.

\vspace*{-5pt}
\section{Proposed method}
\label{sec:method}
\vspace*{-5pt}

In this paper, we propose a method to explore multiple solutions of a linear inverse problem, starting from a first solution $G(\z_0)$. The method exploits geometrical properties of the GAN latent space $\mathcal{Z}$
to navigate in a neighborhood of $\z_0$ in such a way that the new generated data preserve the condition in Eq. \eqref{eq:model} (i.e., they are solutions to the inverse problem) while manifesting novel semantic information with respect to $G(\z_0)$.

The latent space $\mathcal{Z}$ can be seen as a Riemannian manifold, and a GAN $G$ parametrizes a submanifold to the data space $\mathcal{X}$, and, ultimately to the measurement space $\mathcal{Y}$ via the composition of generator and forward model $\phi=G \circ \A$.
Wang and Ponce \cite{wang2021aGANGeom} argue that the geometry of $\mathcal{Z}$ in a neighborhood of $\z_0$ can be approximated by a positive semi-definite quadratic form $H(\z_0)$:
\vspace*{-3pt}
\begin{align*}
d^{2} (\z_0,\z) \approx  \delta \z^{T} \frac{\partial^{2}d^{2} (\z_0,\z)}{\partial\z^{2}}\biggr\rvert_{\z_0} \delta \z, \quad \H(\z_0) := \frac{\partial d^{2} (\z_0,\z)}{\partial\z^{2}}\biggr\rvert_{\z_0},
\end{align*}
\vspace*{-2pt}
dependent on a distance metric $d$ between latent vectors. While the authors in \cite{wang2021aGANGeom} define $d$ between latents as the distance in the generated data space $\mathcal{X}$, we also consider it in the measurement space $\mathcal{Y}$, since we are interested in exploring how variations in the measurement space affect the latent geometry. In particular, we define $d_{\mathcal{Y}}(\z_1,\z_2) := \Vert \phi(\z_1) - \phi(\z_2) \Vert_2^2 = \Vert \A G(\z_1) - \A G(\z_2) \Vert_2^2$ and induce the corresponding manifold described by Riemannian metric:
\begin{align*}\label{eq:hess}
\H_{\mathcal{Y}}(\z_0) = \frac{1}{2} \frac{\partial^{2}}{\partial \z^{2}} \Vert \phi(\z) - \phi(\z_0) \Vert_2^2 \vert_{\z_0} = \J_{\phi}^{T}(\z_0) \J_{\phi}(\z_0),
\end{align*}
being $\J_{\phi}(\z_0) = \partial_{\z}\phi(\z)\vert_{\z_0}$ the Jacobian of $\phi=G \circ \A$ evaluated at point $\z_0$. Similarly, metric $\H_{\mathcal{X}}(\z_0)$ is induced by a suitable distance in the data space. In this work, we will focus on images and, consequently, we use the LPIPS distance (a perceptual metric defined from features extracted by a pretrained network) \cite{zhang2018unreasonable} as $d_{\mathcal{X}}(\z_1,\z_2) := \text{LPIPS}( G(\z_1), G(\z_2) )$. Backpropagation can be used to compute $\H_{\mathcal{Y}}$ and $\H_{\mathcal{X}}$.

Armed with this characterization of the geometry of the latent space, we seek to generate a new latent vector corresponding to a solution as $\z_1 = \z_0 + \eta\d$, i.e., by perturbing $\z_0$ along a direction $\d$ that maximizes perceptual distance in the image space (large $d_{\mathcal{X}}(\z_1,\z_0)$) but minimizes distance in the measurement space (small $d_{\mathcal{Y}}(\z_1,\z_0)$). In other words, we seek to explore the subspace of $\mathcal{Z}$ around $\z_0$ that is ``null'' with respect to the measurements operator but not so with respect to perceptual distance.

One might wonder whether this is possible at all, and, in fact, the answer is affirmative and relies on two main phenomena. The first was observed by Wang and Ponce \cite{wang2021aGANGeom} and it is the \textit{anisotropy} of the space described by $\H_{\mathcal{X}}(\z_0)$, i.e., $\H_{\mathcal{X}}$ is described by a small number of principal components, meaning that there is a large number of directions that have little to no effect on perceptual quality and some significantly changing it\footnote{\cite{wang2021aGANGeom} also note that the space is \textit{homogeneous}, meaning that this property is valid everywhere, regardless of the specific $\z_0$.}. We empirically observe the same regarding the geometry induced by the measurements fidelity, i.e., $\H_{\mathcal{Y}}(\z_0)$. The second phenomenon, which is at the basis of our work, is that the directions from $\H_{\mathcal{X}}$ and $\H_{\mathcal{Y}}$ can be empirically decoupled. This means that it is indeed possible to find directions that significantly affect perceptual distance while having little to no impact on measurements, yielding novel solutions to the inverse problem.

\begin{figure}[t]
\vspace*{-10pt}
\begin{algorithm}[H]
\caption{e-GLASS: exploring GAN LAtent Space Solutions}\label{algo}
\begin{algorithmic}
\Require $\z_0$, $K$
\Ensure New solution $\hat{\x}$
\State Compute Hessian $\H_{\mathcal{Y}}(\z_0) = \frac{1}{2} \frac{\partial^{2}}{\partial \z^{2}} \Vert \A G(\z) - \A G(\z_0) \Vert_2^2$
\State Compute Hessian $\H_{\mathcal{X}}(\z_0) = \frac{1}{2} \frac{\partial^{2}}{\partial \z^{2}} \text{LPIPS}( G(\z),G(\z_0))$
\State Compute eigenvectors $\H_{\mathcal{Y}}(\z_0)=\U \Lambda \U^T$
\State Compute eigenvectors $\H_{\mathcal{X}}(\z_0)=\V \Omega \V^T$
\State $\d \gets \v^{K}$
\State $\mathcal{J}=\lbrace$top-k eigenvectors of $\H_{\mathcal{Y}} \rbrace$ 
\For{$j \in \mathcal{J}$}
\State $\d \gets \d - (\d^{T}\u^{j}) \u^{j}$
\State $\d \gets \d / \Vert \d \Vert$
\EndFor
\State $\hat{\x} = G(\z_0 + \eta \d)$
\end{algorithmic}
\end{algorithm}
\vspace*{-20pt}
\end{figure}

Algorithm \ref{algo} summarizes our proposed e-GLASS scheme to find such directions. We first start by finding the latent code $\z_0$ corresponding to a single solution by means of any state-of-the-art GAN inversion technique. Then we compute the Hessians $\H_{\mathcal{Y}}(\z_0)$, $\H_{\mathcal{X}}(\z_0)$ and their eigenvectors: $\H_{\mathcal{Y}}(\z_0)=\U \Lambda \U^T$, $\H_{\mathcal{X}}(\z_0)=\V \Omega \V^T$. We then need to measure the coupling between the two sets of eigenvectors via the coupling matrix $\C=\U^T\V$. It is expected that the top eigenvectors in $\U$ are coupled with the top eigenvectors in $\V$ as large perceptual distances typically also correspond to large differences on the measurements. However, the bottom eigenvectors are not correlated, indicating that the corresponding null spaces do not necessarily intersect each other. The most interesting directions for our problem are the eigenvectors that are among the top in $\V$ but do not correlate with the top eigenvectors in $\U$. However, directly choosing such direction is in general suboptimal, as it might still increase the distance in the measurement space more than desired. 

To solve this problem, we propose a geometrical method that removes the most relevant correlations with the top eigenvectors of $\H_{\mathcal{Y}}$. This allows to obtain a new direction that is hopefully still creating perceptually significant differences but projected as much as possible onto the null space of $\H_{\mathcal{Y}}$ to minimally change the measurements. To do this, we first choose $\d = \v^{K}$ as the $K$-th top eigenvector to discard the very top eigenvectors that are coupled with the top ones in $\U$. Then, we project $\v^{K}$ onto the hyperplane orthogonal to the top eigenvectors $\u^{j}$ with correlation larger than a threshold:
\vspace*{-5pt}
\begin{align*}
\d \leftarrow \d - (\d^{T}\u^{j}) \u^{j}, \quad \d \leftarrow \d / \Vert \d \Vert  .
\end{align*}
This procedure is iterated until the resulting direction has no significant correlation to the top eigenvectors of $\H_{\mathcal{Y}}$. This leads to a projection of $\v^{K}$ onto the null space of $\H_{\mathcal{Y}}$. Multiple solutions to the inverse problem can be explored by either changing the step $\eta$ along the direction, or trying a new direction by computing $\d$ starting from $\v^{K-1}$, $\v^{K-2}$, ...

\vspace*{-5pt}
\section{Experimental Results}
\label{sec:results}
\vspace*{-5pt}

In this section, we experimentally evaluate the proposed method against state-of-the-art techniques to explore multiple solutions. While the proposed method is general and holds for different generative models and different inverse problems, we focus on two notable inverse problems, i.e., image super-resolution (SR) and inpainting (IP), presenting results for two different generative models, i.e. BigGAN \cite{brock2018large} and PGGAN \cite{karras2017progressive}.
For super-resolution, we downscale the $256 \times 256$ image to a $32 \times 32$ image, while for inpainting, we delete a semantically interesting area of a face from a full image of size $1024 \times 1024$. 

\begin{figure}
\centering
\begin{subfigure}{0.49\columnwidth}
    \includegraphics[width=\textwidth]{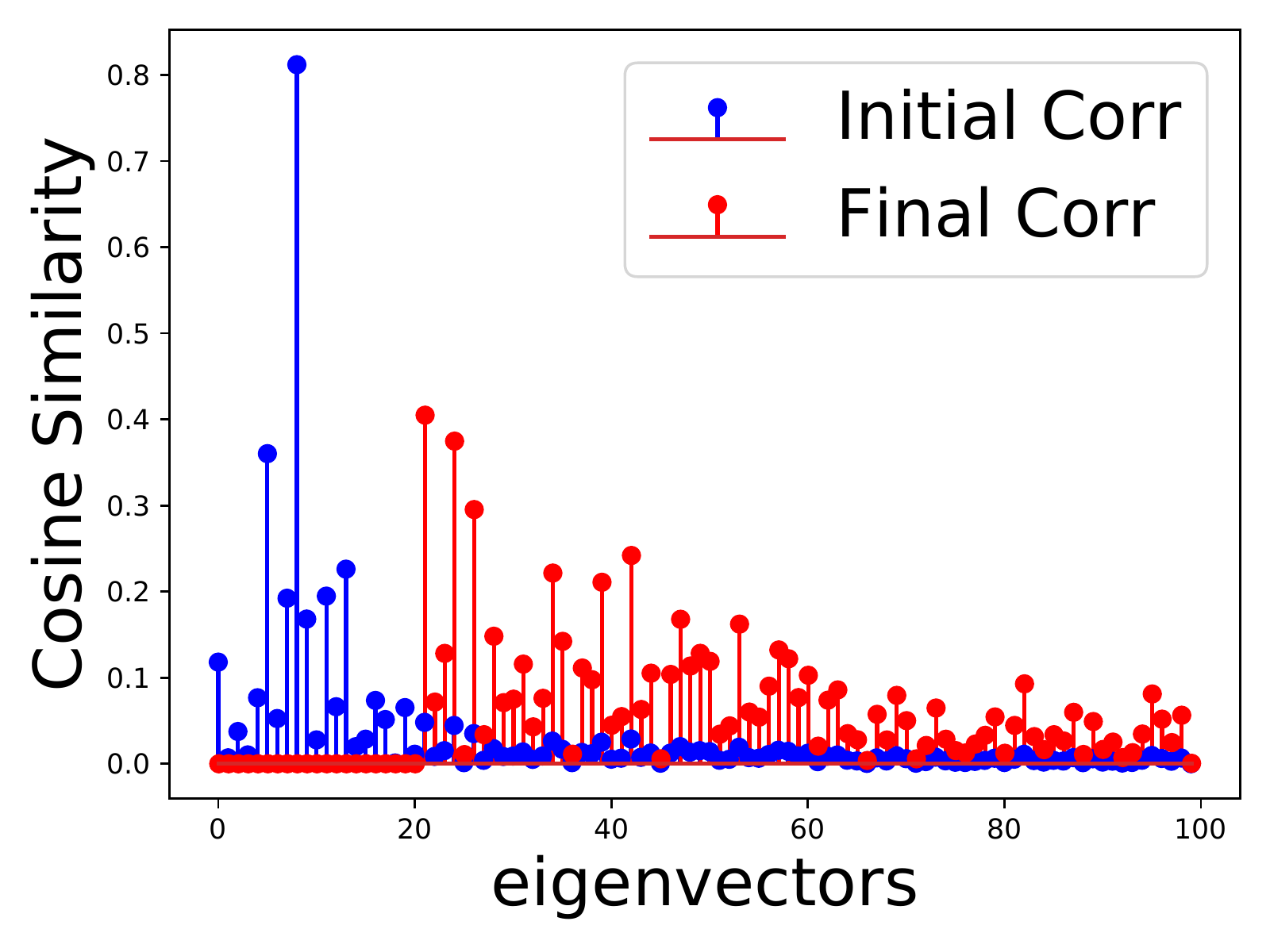}
    \caption{Correlation with $\U$.}
\end{subfigure}
\begin{subfigure}{0.49\columnwidth}
    \includegraphics[width=\textwidth]{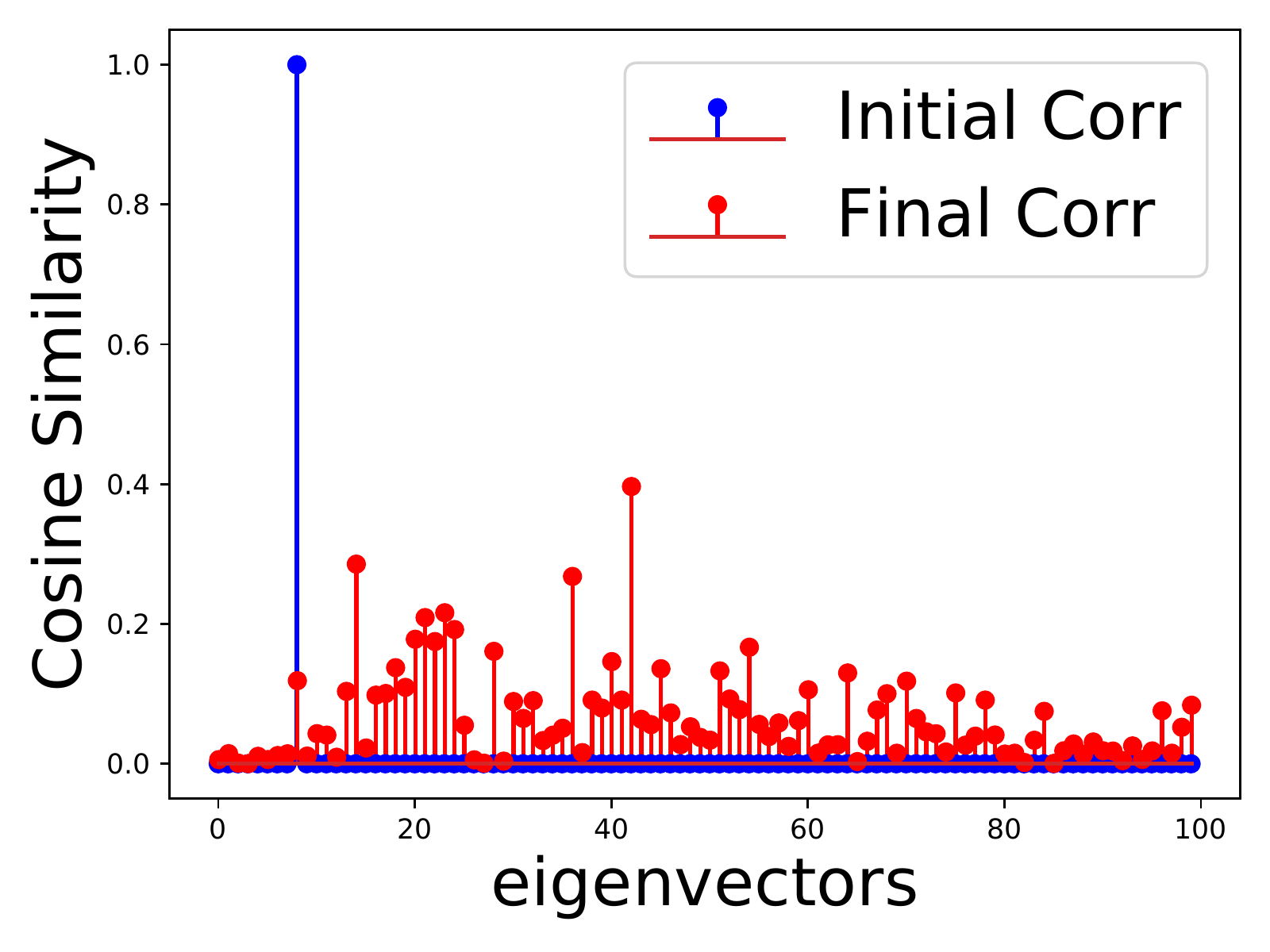}
    \caption{Correlation with $\V$.}
\end{subfigure}
\vspace*{-5pt}
\caption{Correlation of initial direction $\v^K$ and final direction $\d$ with eigenvectors of latent space metrics $\H_\mathcal{Y}$ and $\H_\mathcal{X}$. The final direction is orthogonal to directions inducing large change in measurements, but correlates with directions inducing significant perceptual change.}
\label{fig:corr}
\vspace*{-10pt}
\end{figure}

We first present empirical evidence about our claims in Sec.\ref{sec:method} that directions that induce little change on measurements and significant perceptual change on the reconstructions do exist. To show this we want to see that the chosen direction correlates with the top eigenvectors of $\V$ while being as orthogonal as possible to the top eigenvectors of $\U$. Fig. \ref{fig:corr} shows in blue the correlation coefficient between the starting direction $\v^K$ and the eigenvectors in $\U$ and $\V$ and in red the same correlations but with respect to the final direction $\d$ provided by our method. It can be noticed that the final direction has been successfully orthogonalized with respect to the directions inducing significant variation on the measurements, while it retains good correlation with directions inducing perceptual change.

We now qualitatively and quantitatively examine the performance of the proposed method for the chosen inverse problems in comparison with PULSE \cite{menon2020pulse}. PULSE generates multiple solutions by solving the GAN inversion problem multiple times from different random initialization, in the hope of converging to a different local minimum. For our proposed method, we generate the initial solution by means of the state-of-the-art GAN inversion technique proposed by Abu Hussein et al. \cite{hussein2020image}, where the inversion problem in Eq. \eqref{eq:stageI_loss} also optimizes with respect to the GAN parameters to finetune them. Once the initial solution is computed, we apply Algorithm 1 to find a new solution to the problem.

\begin{figure*}
\centering
\begin{subfigure}[b]{0.16\textwidth}
\includegraphics[width=\textwidth]{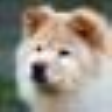}
\end{subfigure}
\begin{subfigure}[b]{0.16\textwidth}
\includegraphics[width=\textwidth]{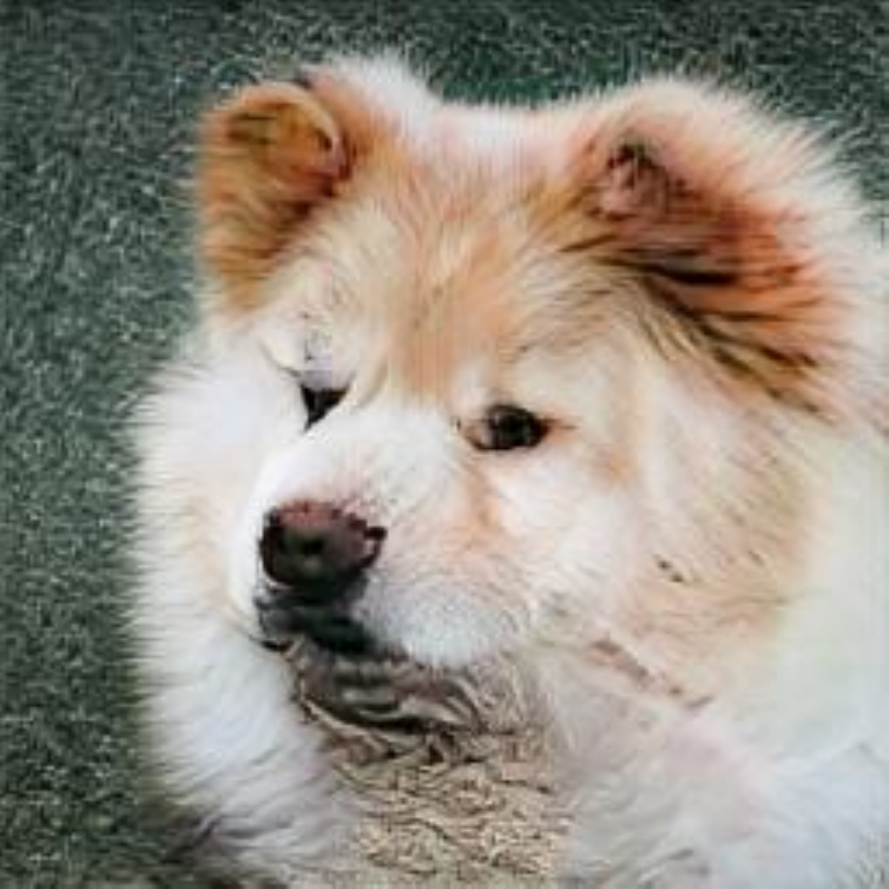}
\end{subfigure}
\begin{subfigure}[b]{0.16\textwidth}
\includegraphics[width=\textwidth]{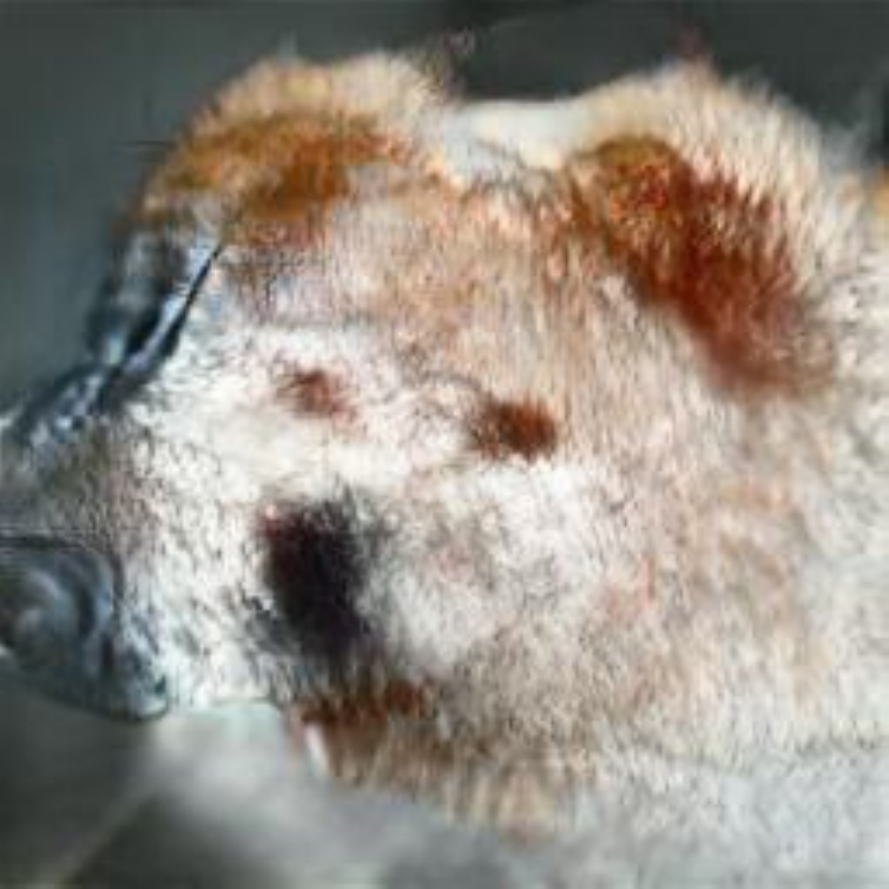}
\end{subfigure}
\begin{subfigure}[b]{0.16\textwidth}
\includegraphics[width=\textwidth]{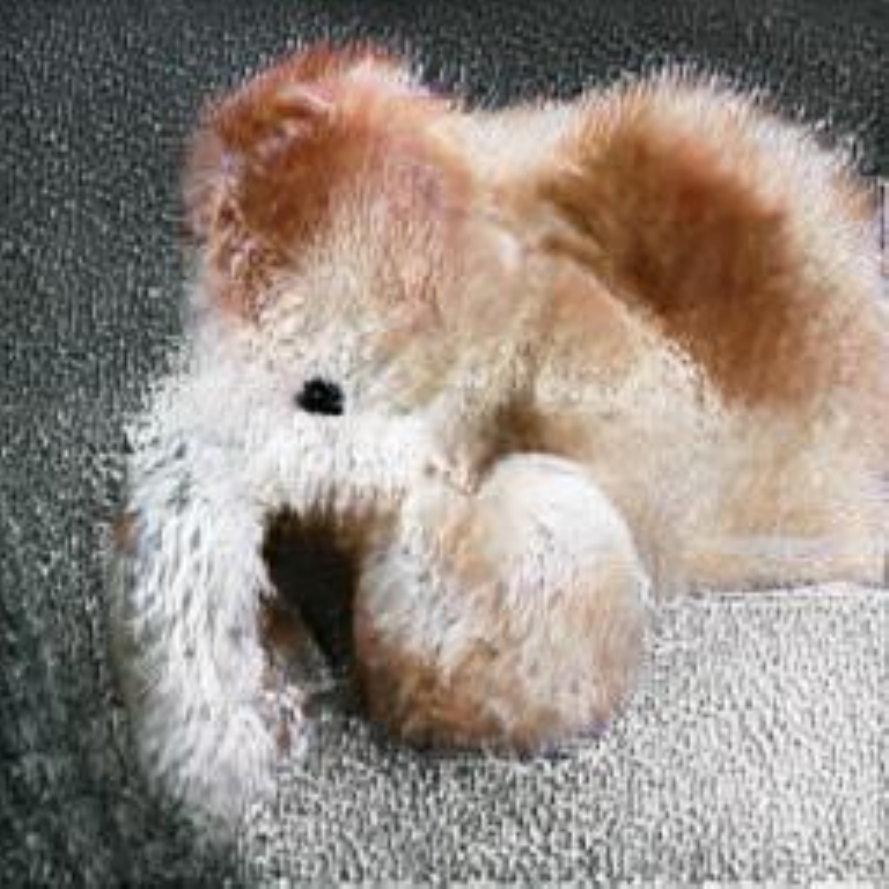}
\end{subfigure}
\begin{subfigure}[b]{0.16\textwidth}
\includegraphics[width=\textwidth]{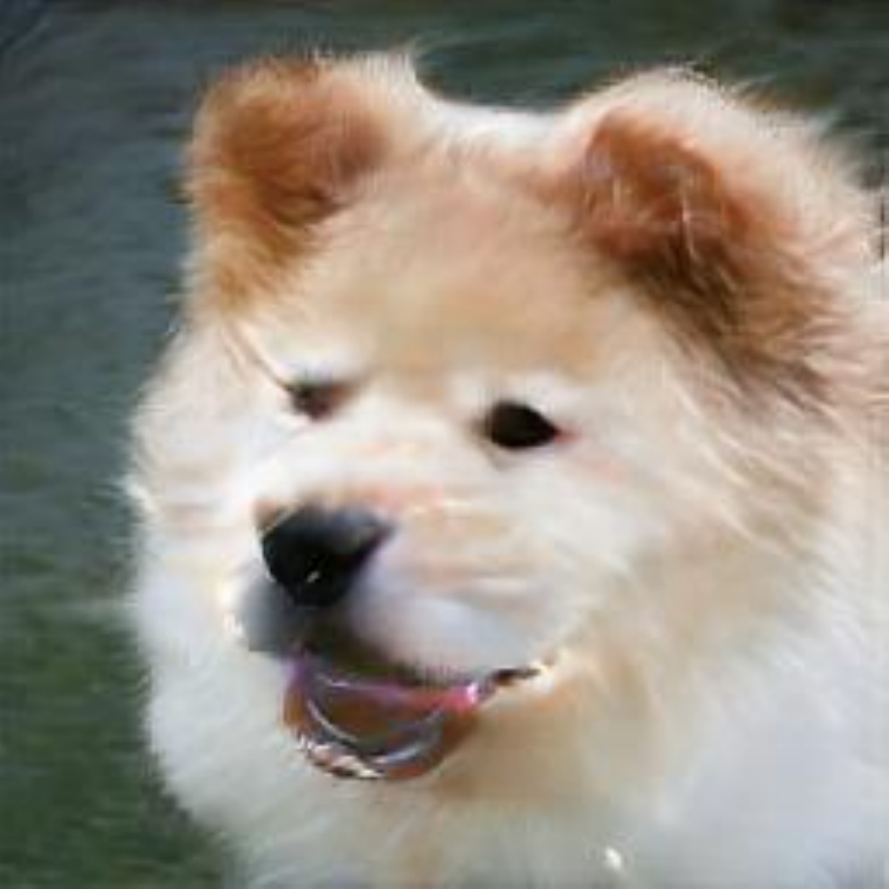}
\end{subfigure}
\begin{subfigure}[b]{0.16\textwidth}
\includegraphics[width=\textwidth]{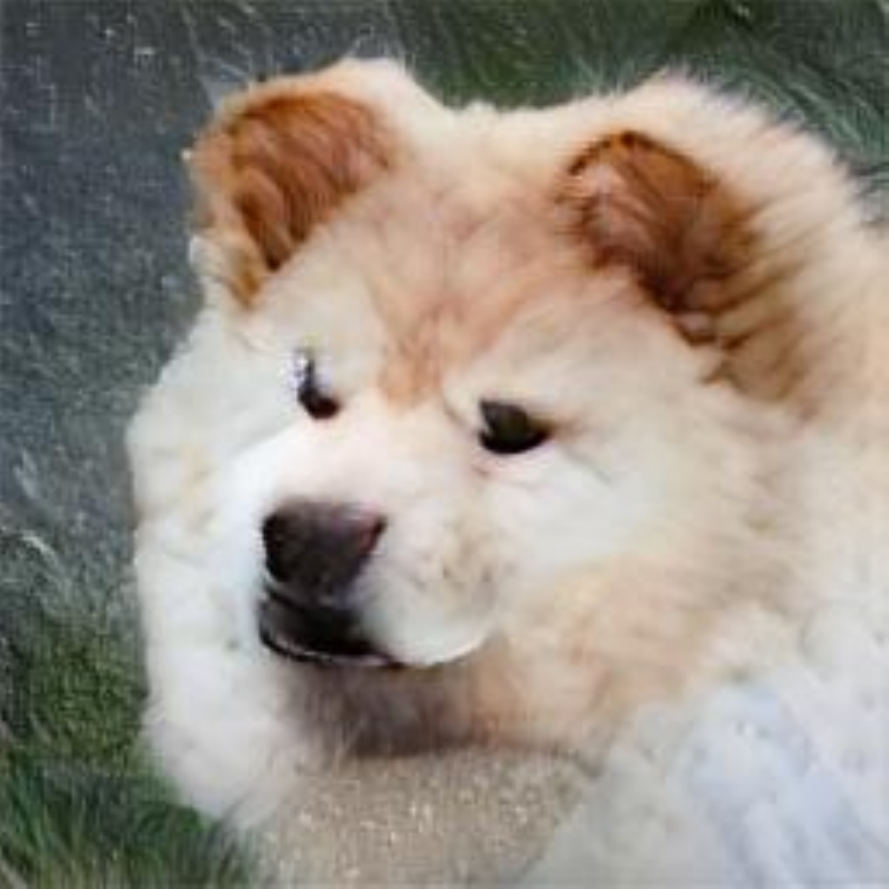}
\end{subfigure}
\\[3pt]
\begin{subfigure}[b]{0.16\textwidth}
\includegraphics[width=\textwidth]{figures/chow_input.pdf}
\end{subfigure}
\begin{subfigure}[b]{0.16\textwidth}
\includegraphics[width=\textwidth]{figures/chow_sol0.pdf}
\end{subfigure}
\begin{subfigure}[b]{0.16\textwidth}
\includegraphics[width=\textwidth]{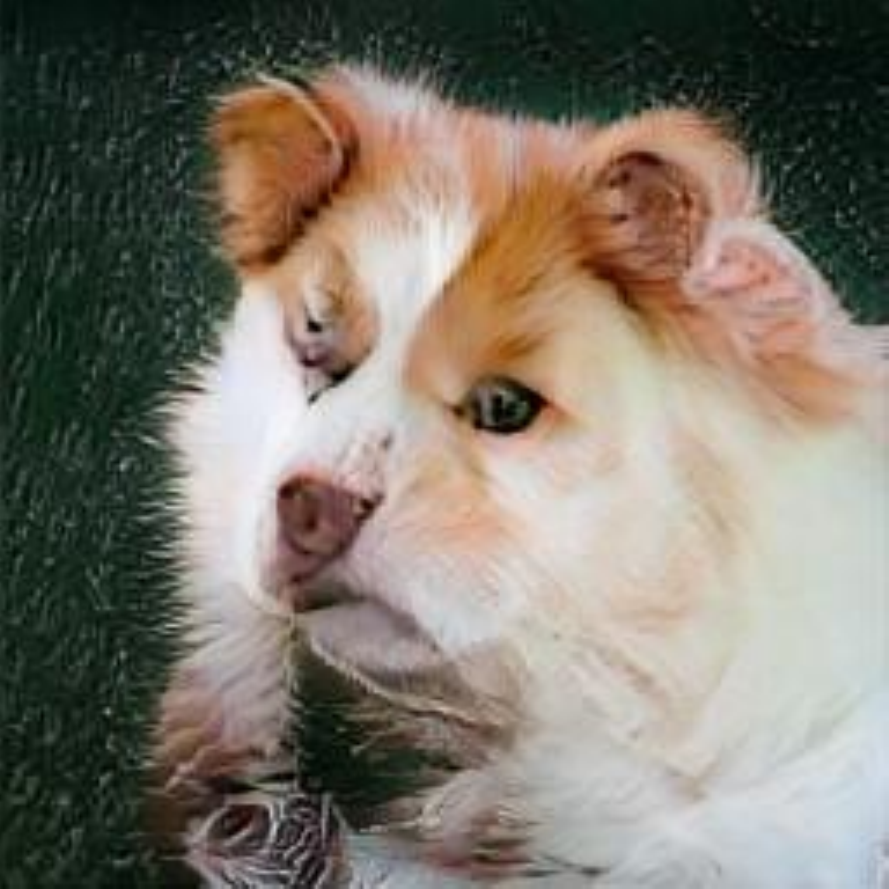}
\end{subfigure}
\begin{subfigure}[b]{0.16\textwidth}
\includegraphics[width=\textwidth]{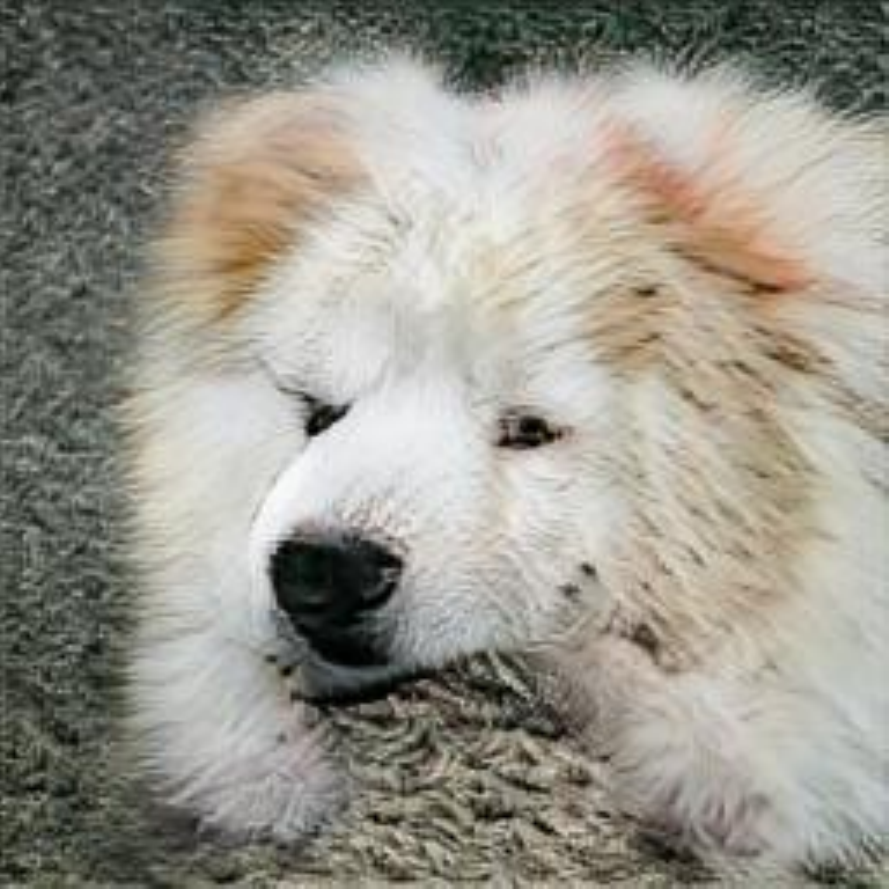}
\end{subfigure}
\begin{subfigure}[b]{0.16\textwidth}
\includegraphics[width=\textwidth]{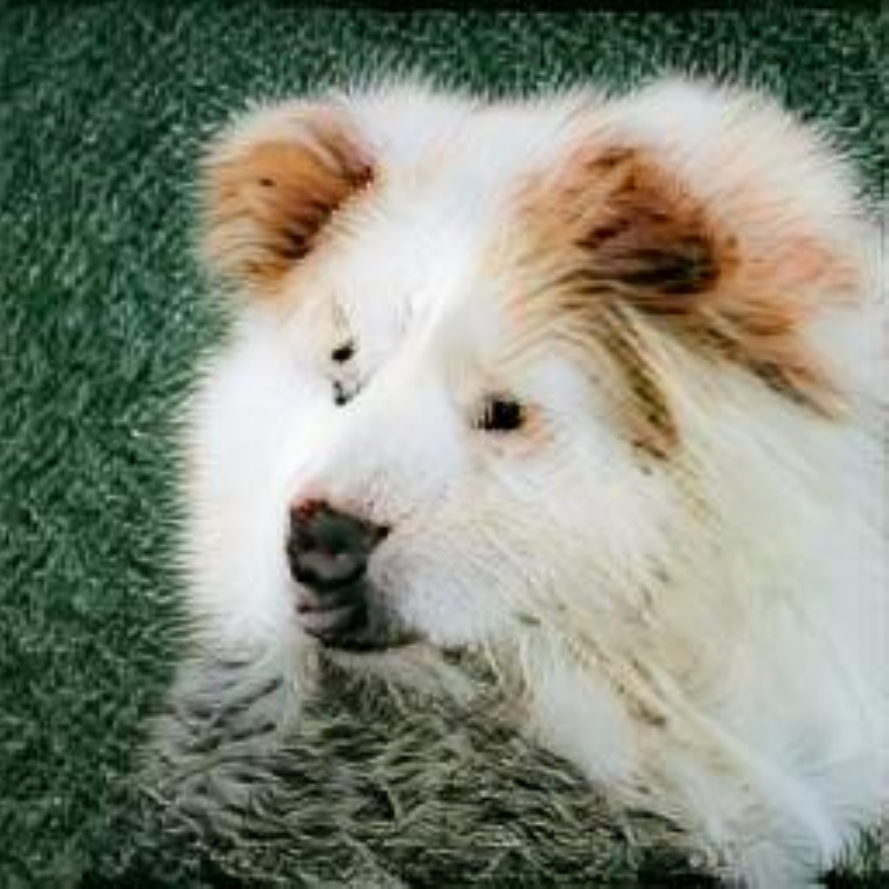}
\end{subfigure}
\begin{subfigure}[b]{0.16\textwidth}
\includegraphics[width=\textwidth]{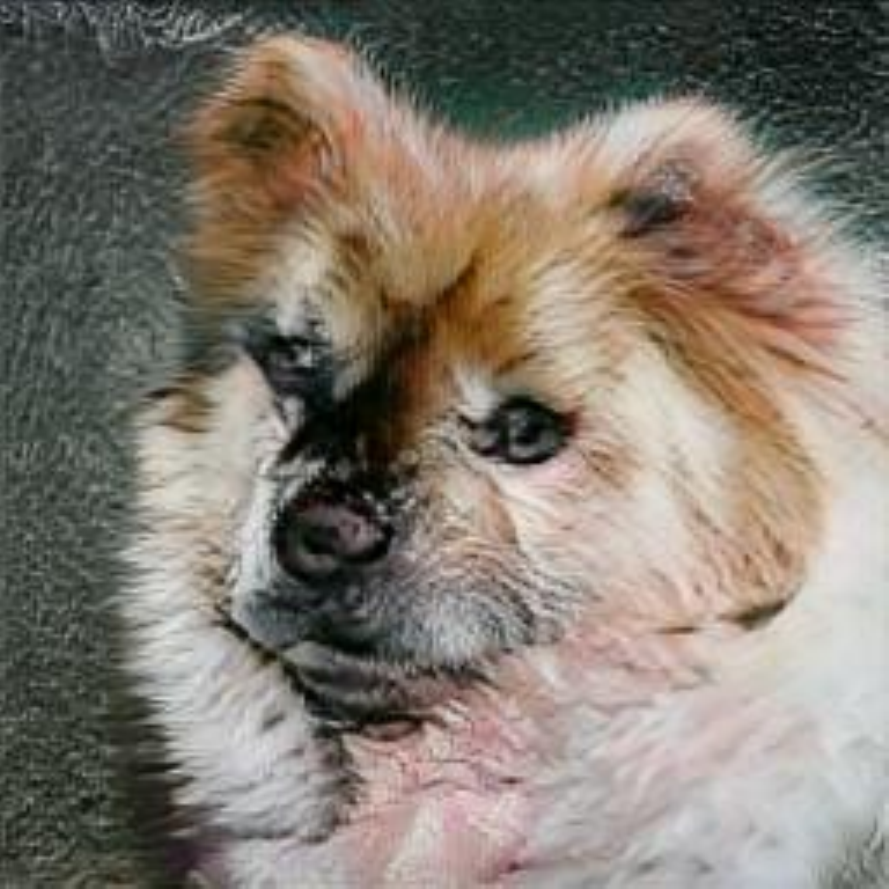}
\end{subfigure}
\\[3pt]
\begin{subfigure}[b]{0.16\textwidth}
\includegraphics[width=\textwidth]{figures/chow_input.pdf}
\end{subfigure}
\begin{subfigure}[b]{0.16\textwidth}
\includegraphics[width=\textwidth]{figures/chow_sol0.pdf}
\end{subfigure}
\begin{subfigure}[b]{0.16\textwidth}
\includegraphics[width=\textwidth]{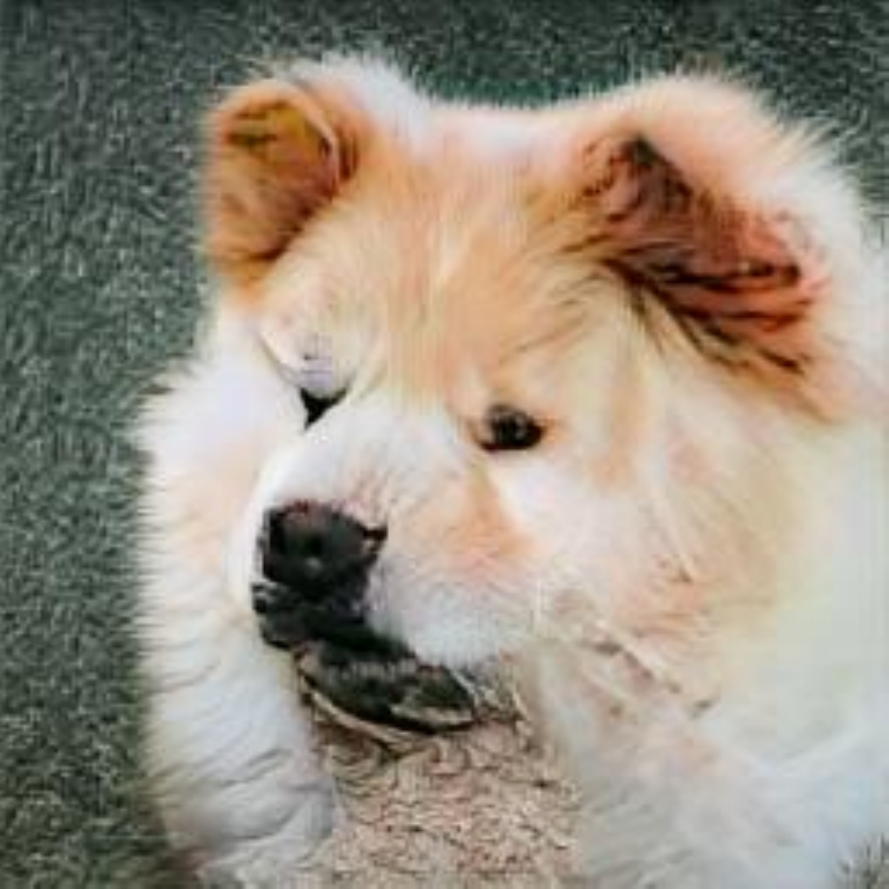}
\end{subfigure}
\begin{subfigure}[b]{0.16\textwidth}
\includegraphics[width=\textwidth]{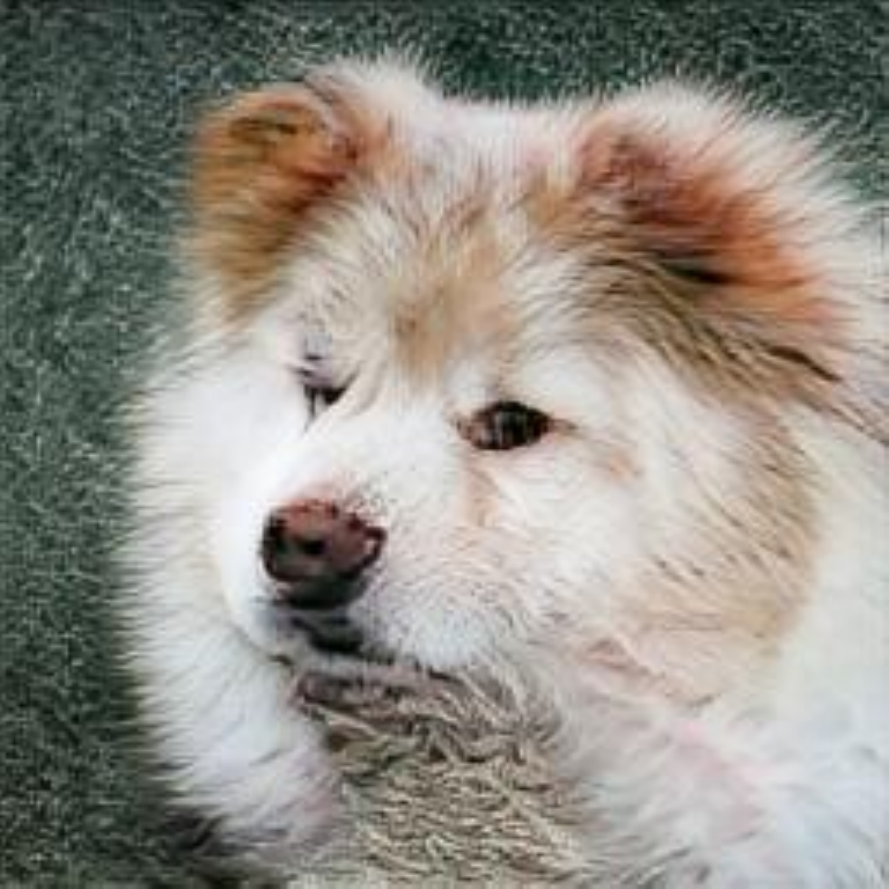}
\end{subfigure}
\begin{subfigure}[b]{0.16\textwidth}
\includegraphics[width=\textwidth]{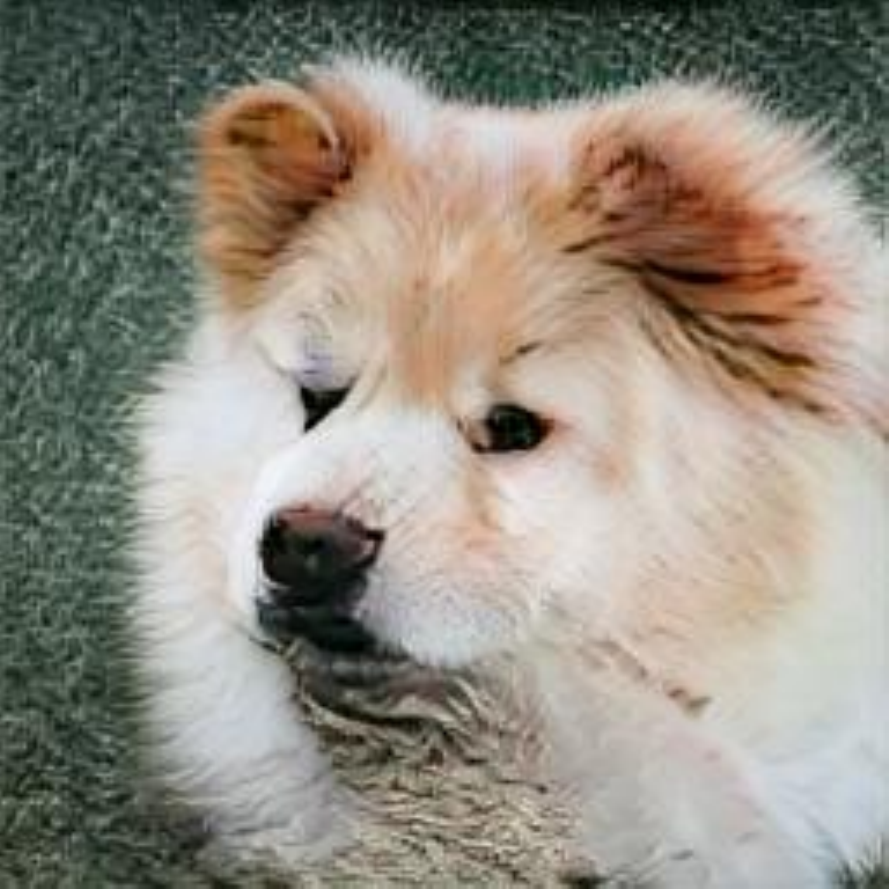}
\end{subfigure}
\begin{subfigure}[b]{0.16\textwidth}
\includegraphics[width=\textwidth]{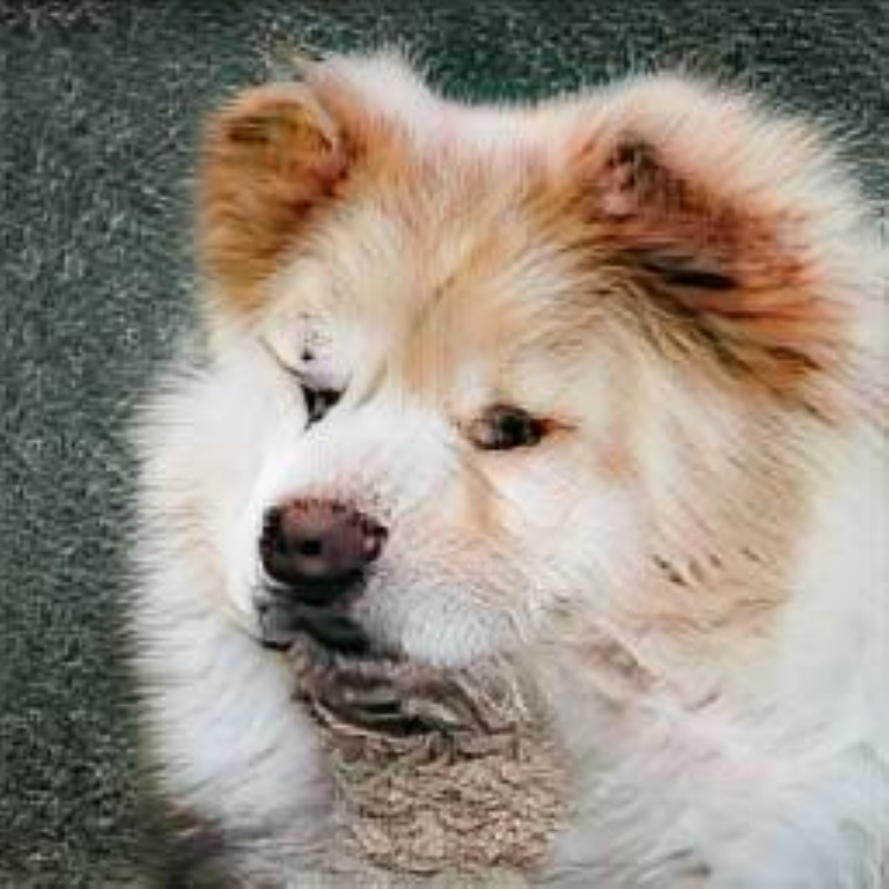}
\end{subfigure}

\caption{Top row: solutions found by PULSE (LR $\ell_2$ range: $[1.8\times10^{-3},3\times10^{-3}]$). Mid row: solutions found by using $\v^8$ and $\v^12$ as directions (LR $\ell_2$ range: $[2.4\times10^{-2},4.5\times10^{-2}]$). Bottom row: solutions found by optimized $\d$ as direction (LR $\ell_2$ range: $[2.9\times10^{-3},4.8\times10^{-3}]$).}
\label{fig:SR_solutions}
\end{figure*}

Fig. \ref{fig:SR_solutions} shows a few results on the SR problem. The middle row shows what reconstructions would be obtained if direction $\v^K$ were used without our proposed algorithm. It can be noticed that there is significant perceptual change but the $\ell_2$ norm with respect to the measurements is poorly constrained, so that these reconstructions can be hardly called feasible solutions. The last row shows the images generated by the direction found by our method. We successfully constrain the $\ell_2$ norm with respect to the measurements below the $10^{-2}$ threshold we consider acceptable for feasibility. At the same time, perceptual variations are still present in those regions where the highly downsampled nature of the measurements leaves more freedom to fill in information, such as the color and shape of the dog's coat (from pale yellow to white, and the shape of the ears). Finally, the top row shows some good solutions found by PULSE. Those solutions are feasible according to our $\ell_2$ criterion but are less perceptually convincing. Indeed while some solutions show different dogs, these present some artifacts around the dog's mouth or some blurring over the whole dog's coat. Instead, the two central solutions in Fig \ref{fig:SR_solutions} (top row), show unnatural dogs found by the PULSE algorithm that still satisfy the $\ell_2$ criterion.

\begin{figure*}
\centering
\begin{subfigure}[b]{0.16\textwidth}
\includegraphics[width=\textwidth]{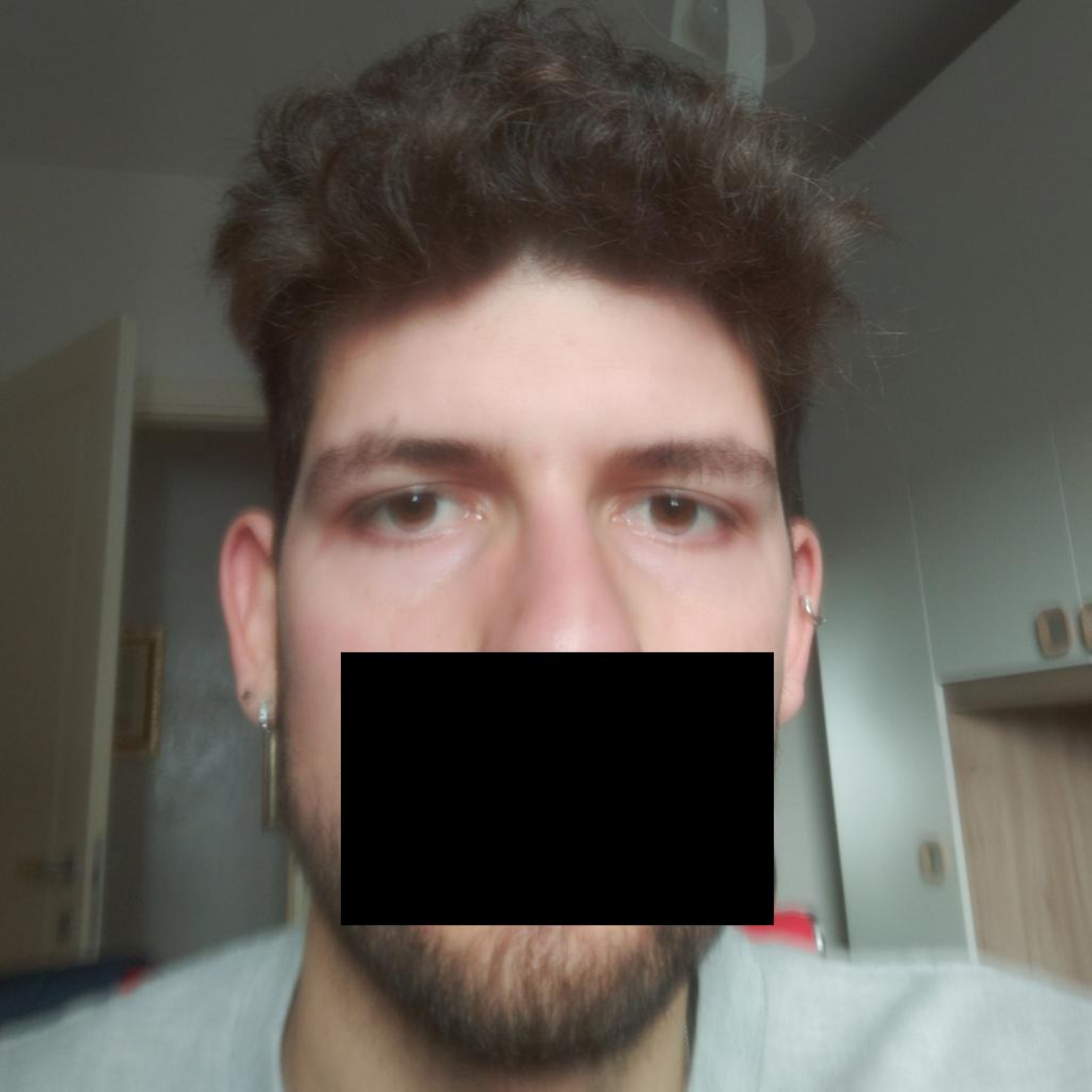}
\end{subfigure}
\begin{subfigure}[b]{0.16\textwidth}
\includegraphics[width=\textwidth]{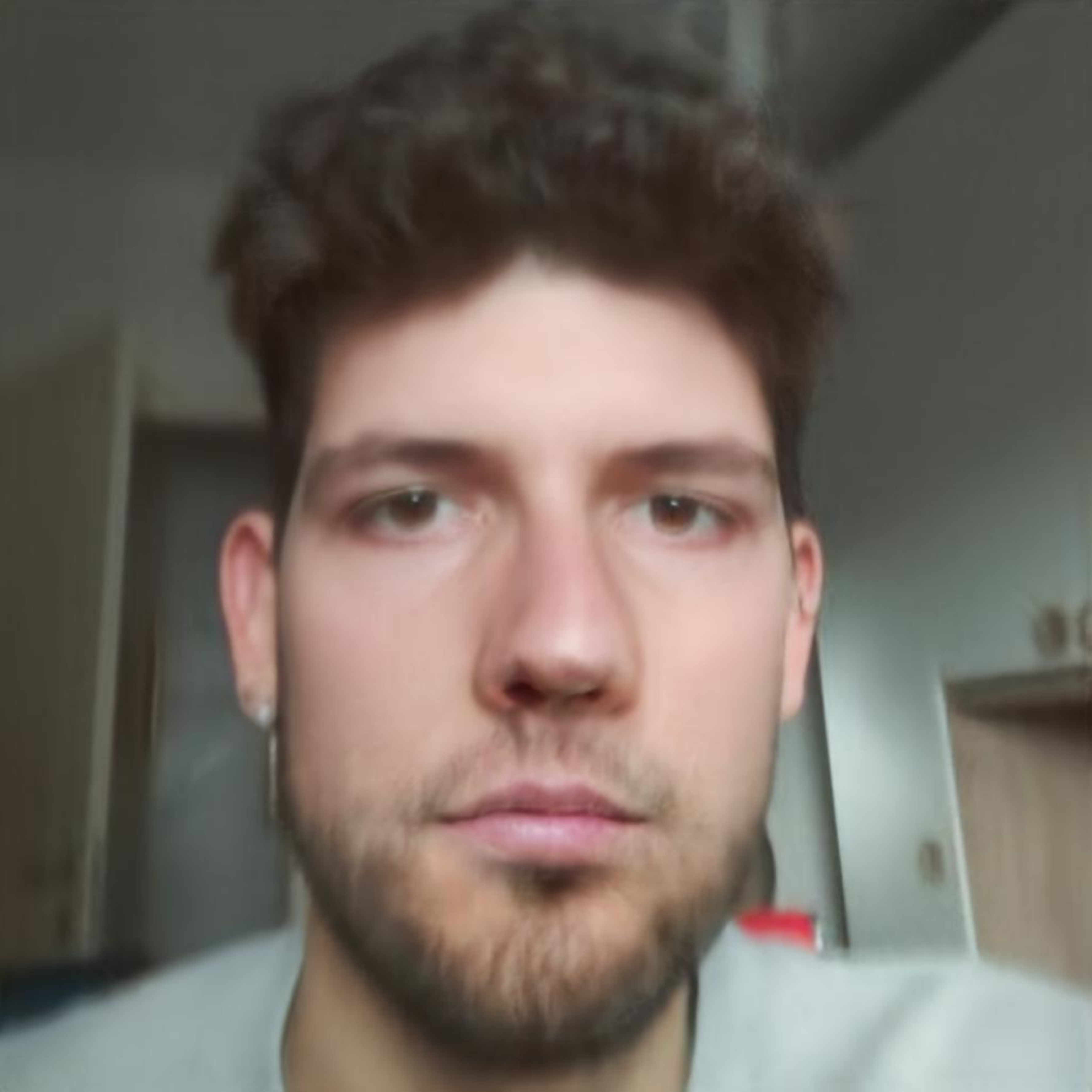}
\end{subfigure}
\begin{subfigure}[b]{0.16\textwidth}
\includegraphics[width=\textwidth]{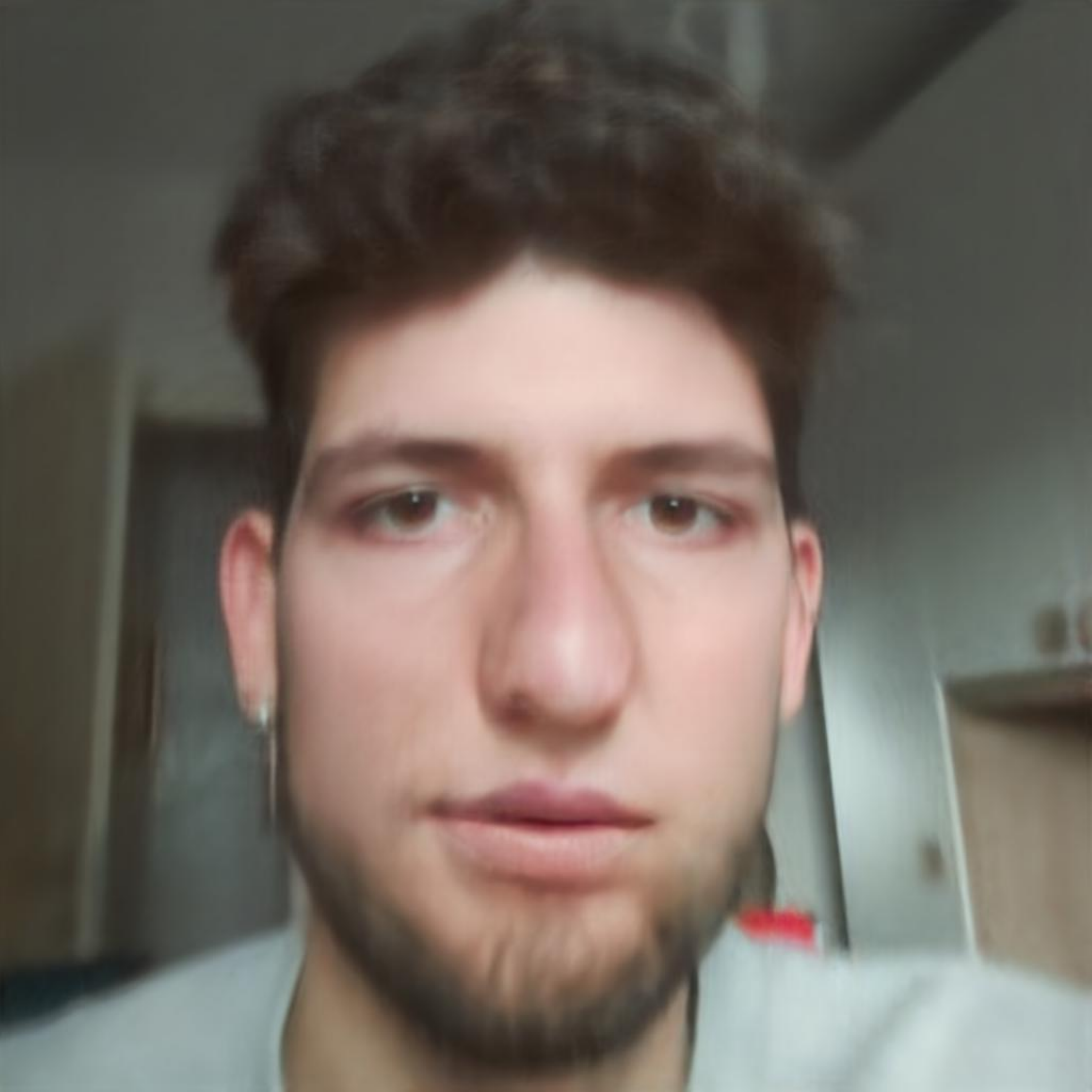}
\end{subfigure}
\begin{subfigure}[b]{0.16\textwidth}
\includegraphics[width=\textwidth]{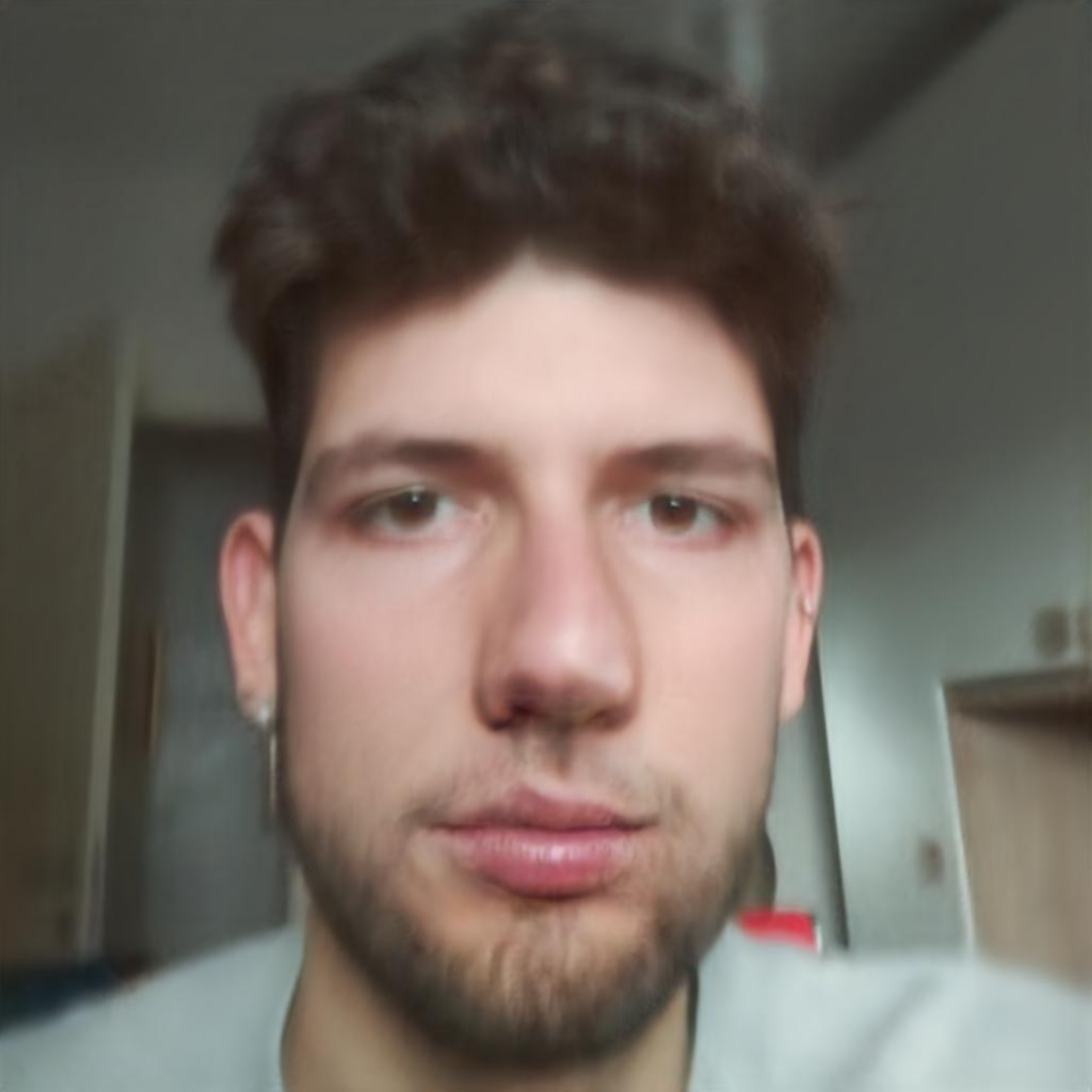}
\end{subfigure}
\begin{subfigure}[b]{0.16\textwidth}
\includegraphics[width=\textwidth]{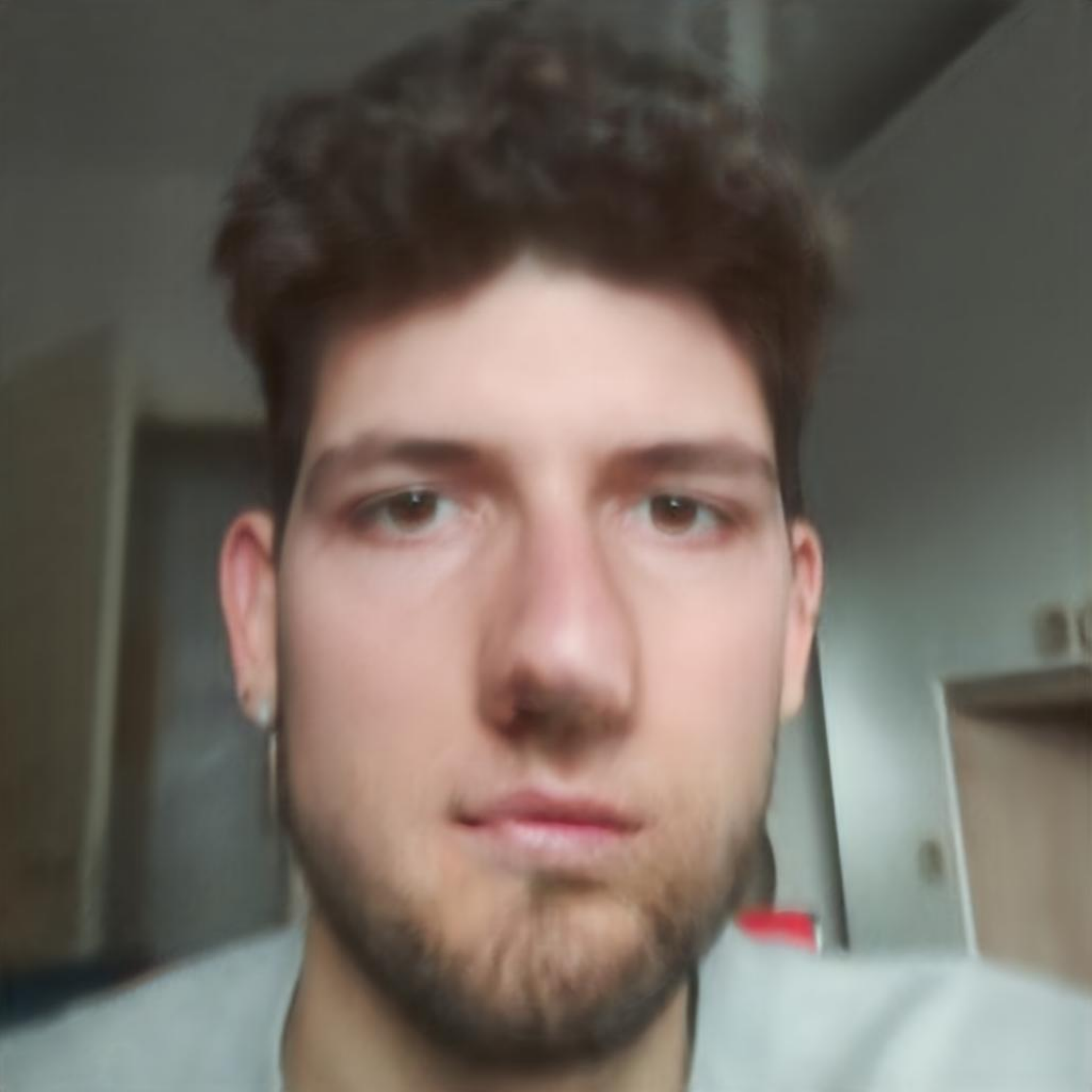}
\end{subfigure}
\begin{subfigure}[b]{0.16\textwidth}
\includegraphics[width=\textwidth]{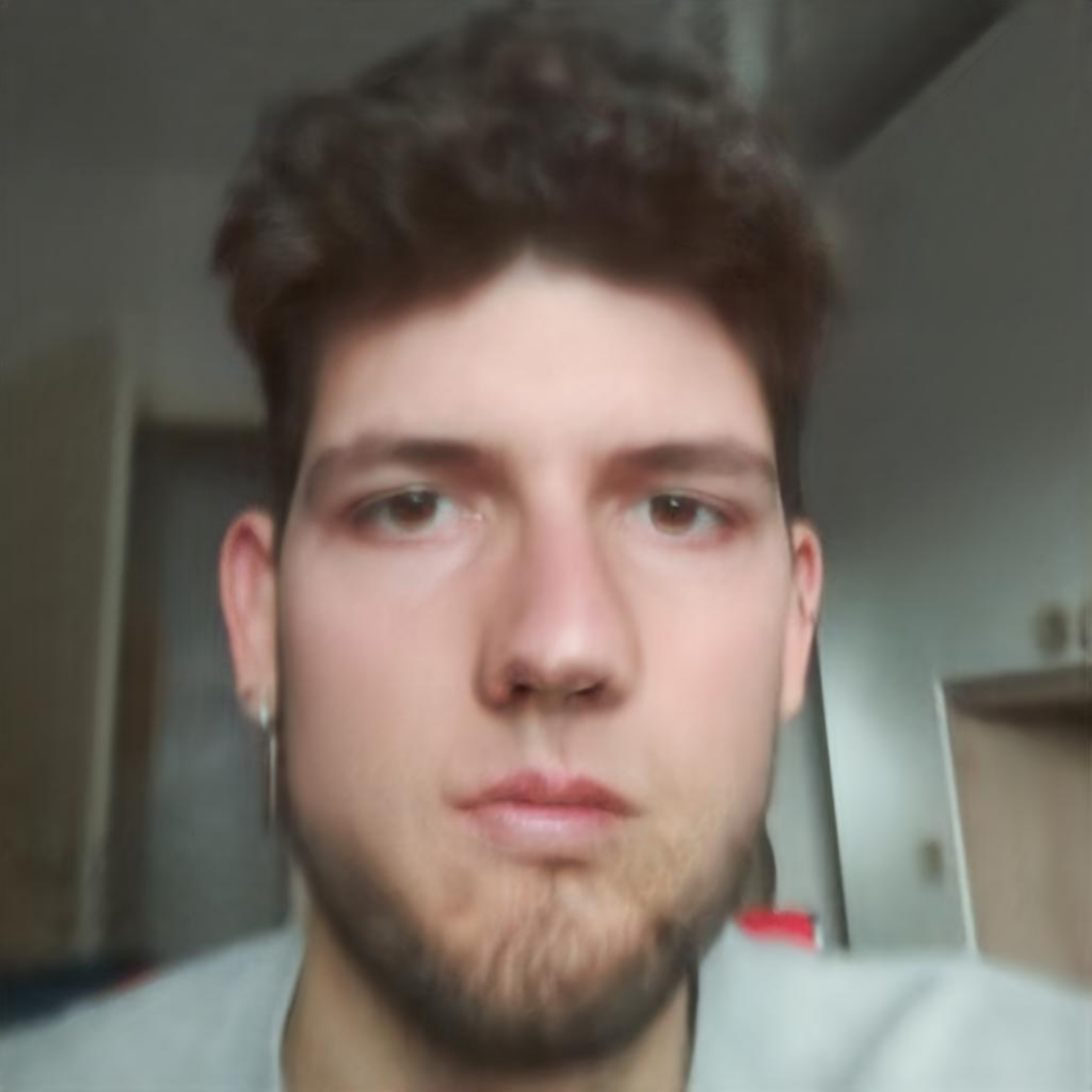}
\end{subfigure}
\\[3pt]
\begin{subfigure}[b]{0.16\textwidth}
\includegraphics[width=\textwidth]{figures/io_input.pdf}
\end{subfigure}
\begin{subfigure}[b]{0.16\textwidth}
\includegraphics[width=\textwidth]{figures/io_sol0.pdf}
\end{subfigure}
\begin{subfigure}[b]{0.16\textwidth}
\includegraphics[width=\textwidth]{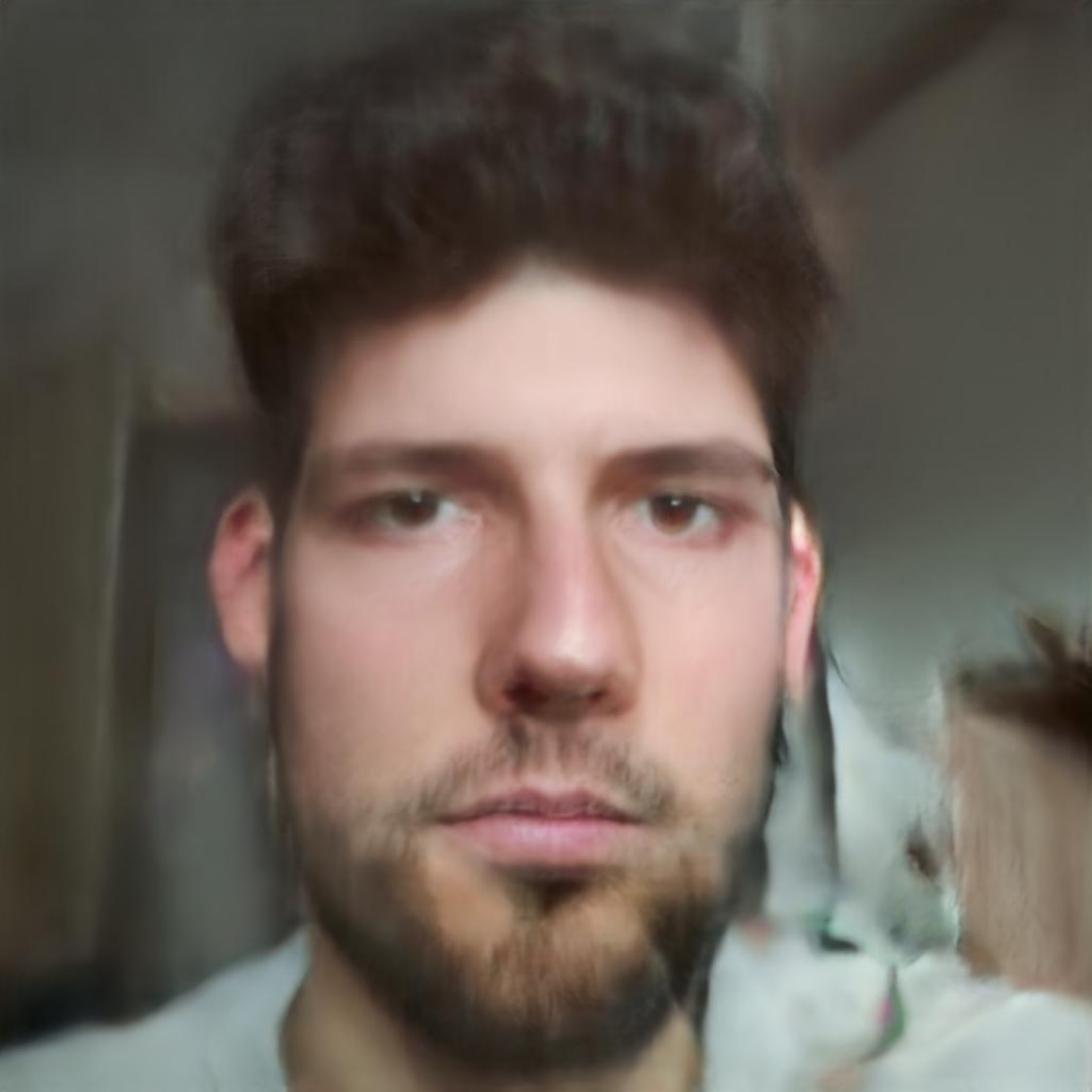}
\end{subfigure}
\begin{subfigure}[b]{0.16\textwidth}
\includegraphics[width=\textwidth]{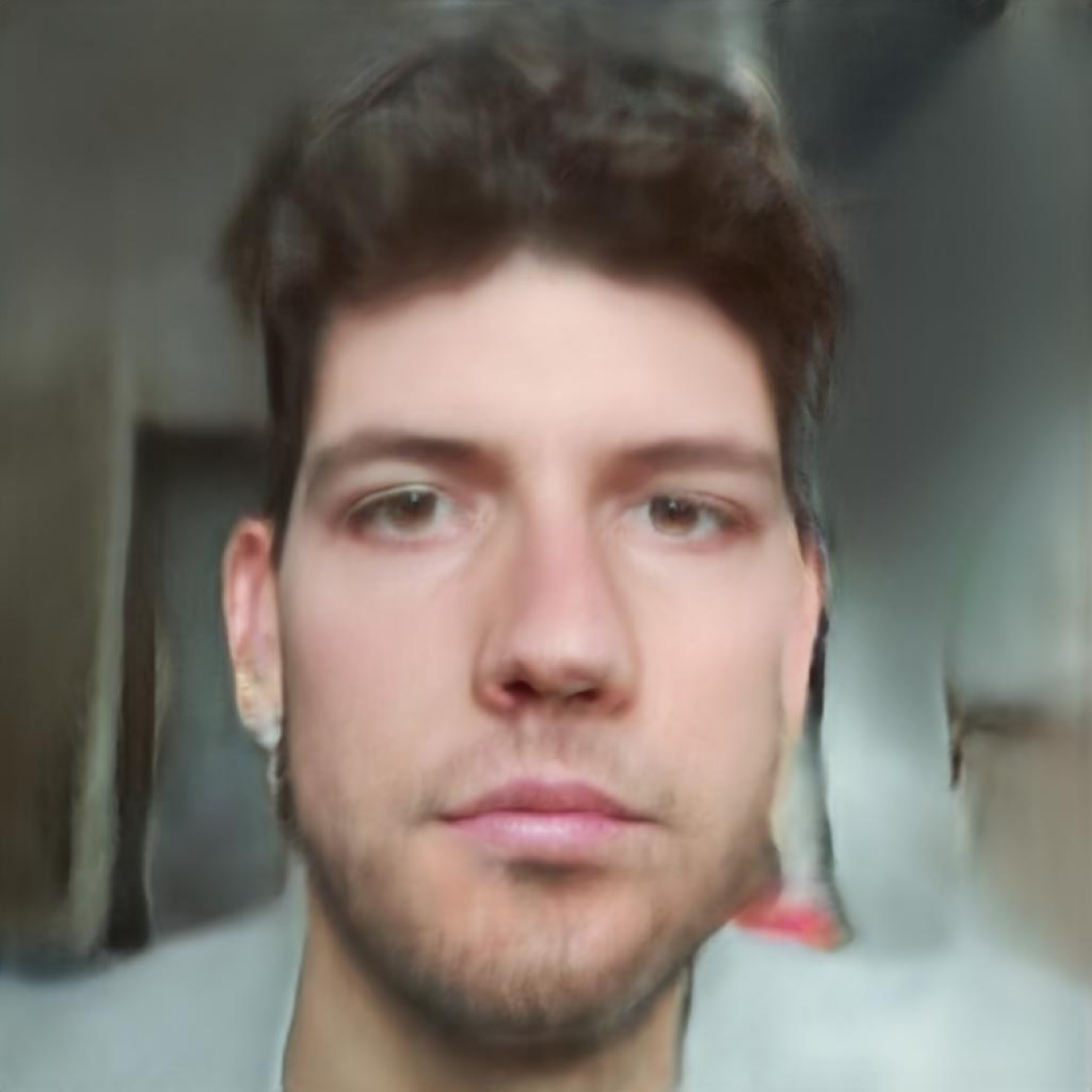}
\end{subfigure}
\begin{subfigure}[b]{0.16\textwidth}
\includegraphics[width=\textwidth]{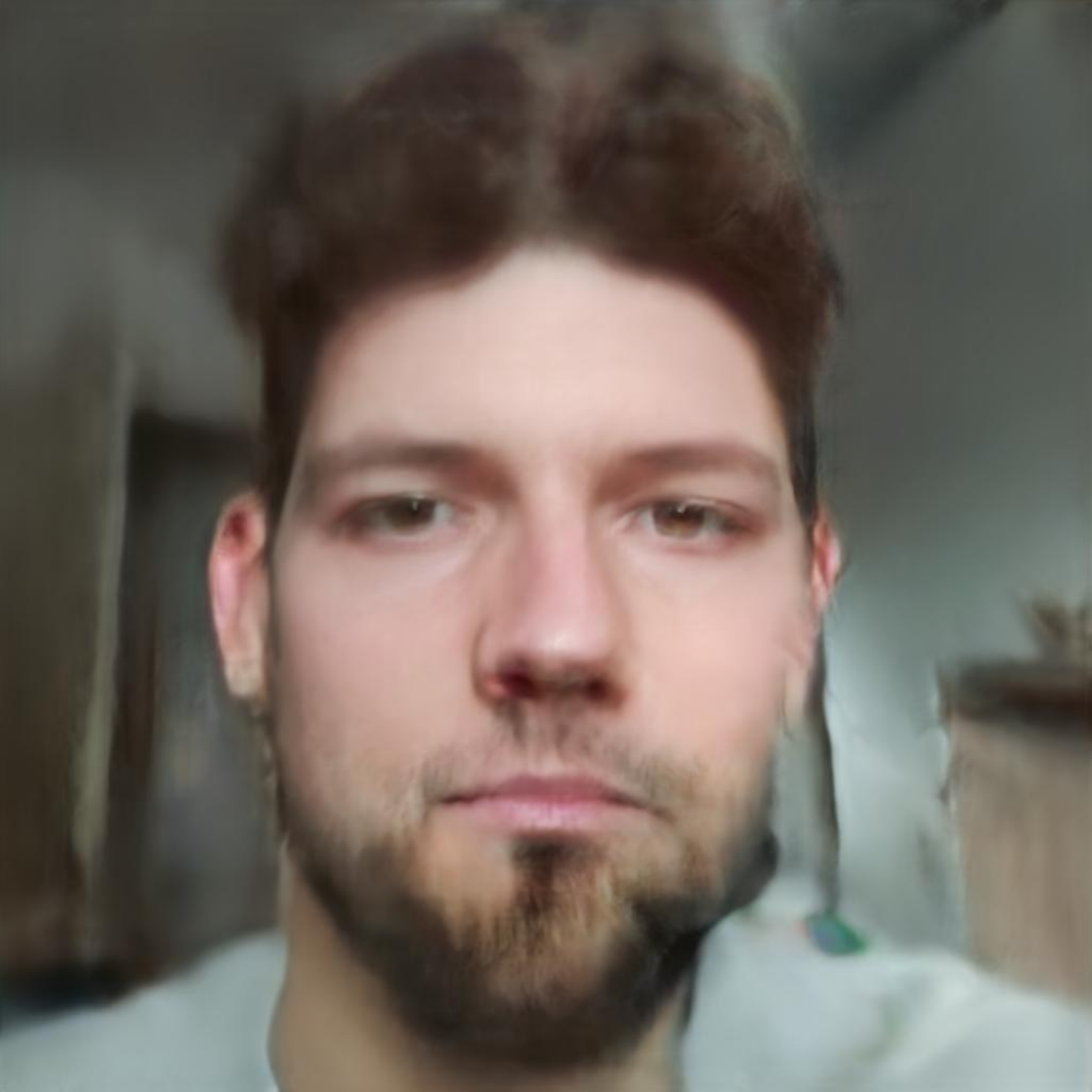}
\end{subfigure}
\begin{subfigure}[b]{0.16\textwidth}
\includegraphics[width=\textwidth]{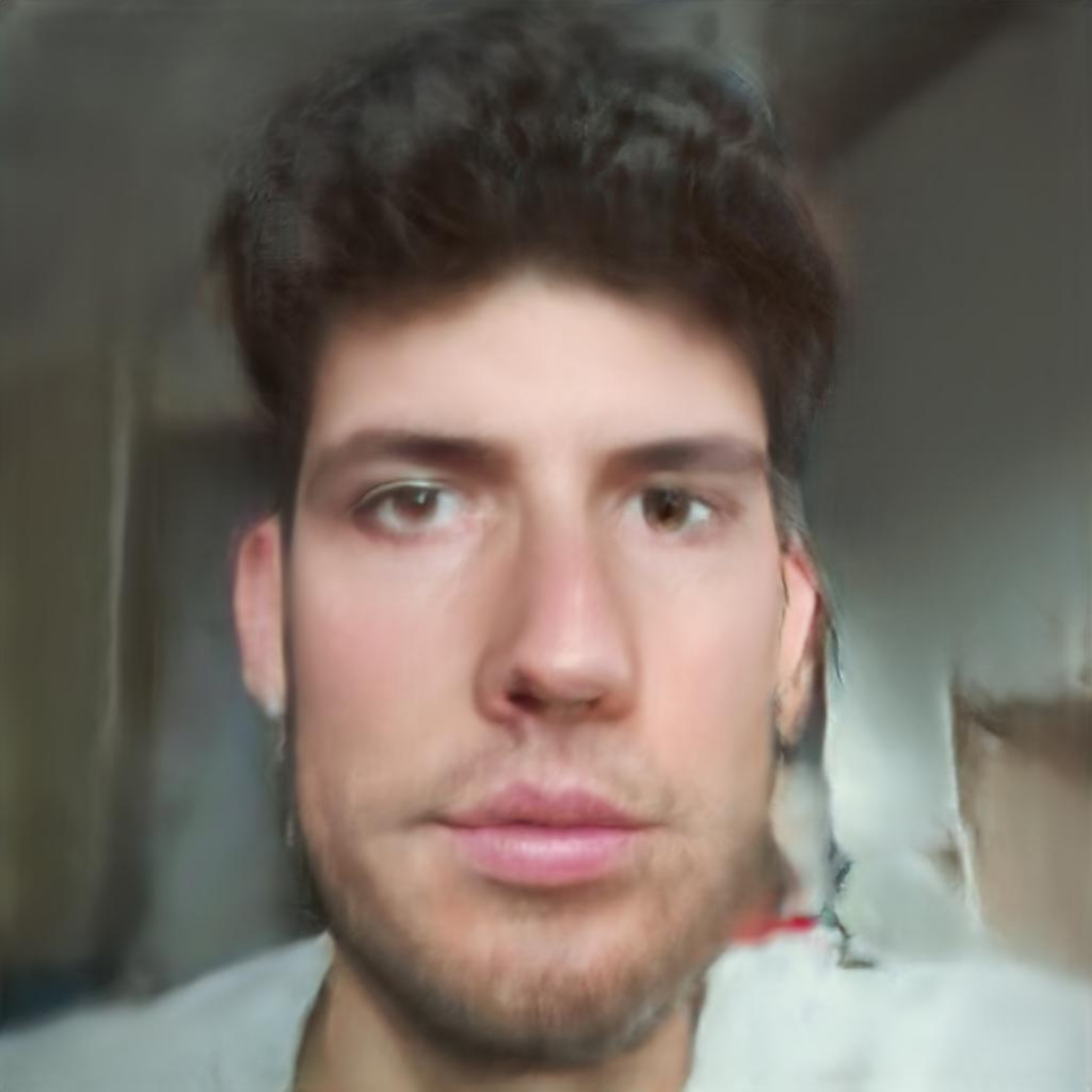}
\end{subfigure}
\\[3pt]
\begin{subfigure}[b]{0.16\textwidth}
\includegraphics[width=\textwidth]{figures/io_input.pdf}
\end{subfigure}
\begin{subfigure}[b]{0.16\textwidth}
\includegraphics[width=\textwidth]{figures/io_sol0.pdf}
\end{subfigure}
\begin{subfigure}[b]{0.16\textwidth}
\includegraphics[width=\textwidth]{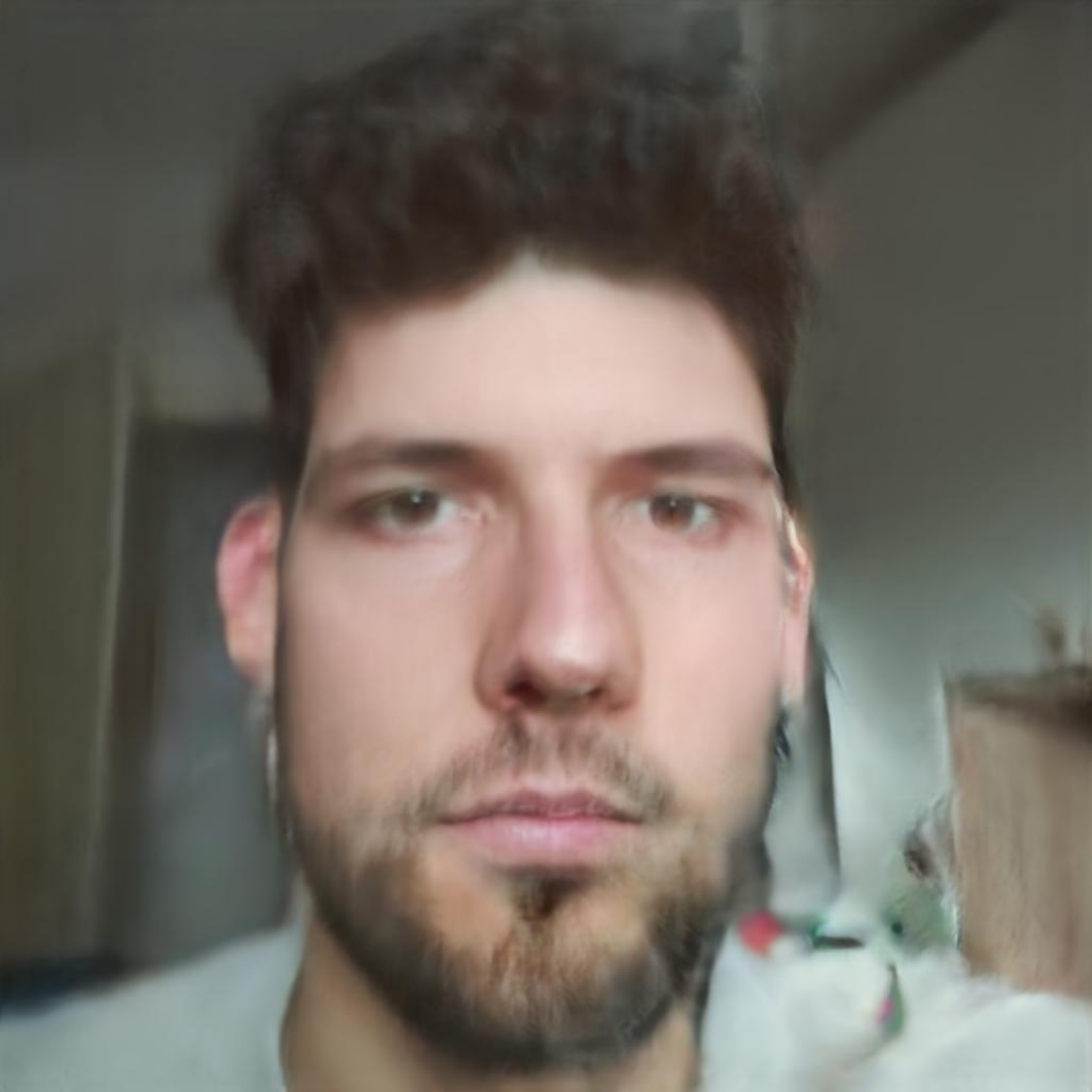}
\end{subfigure}
\begin{subfigure}[b]{0.16\textwidth}
\includegraphics[width=\textwidth]{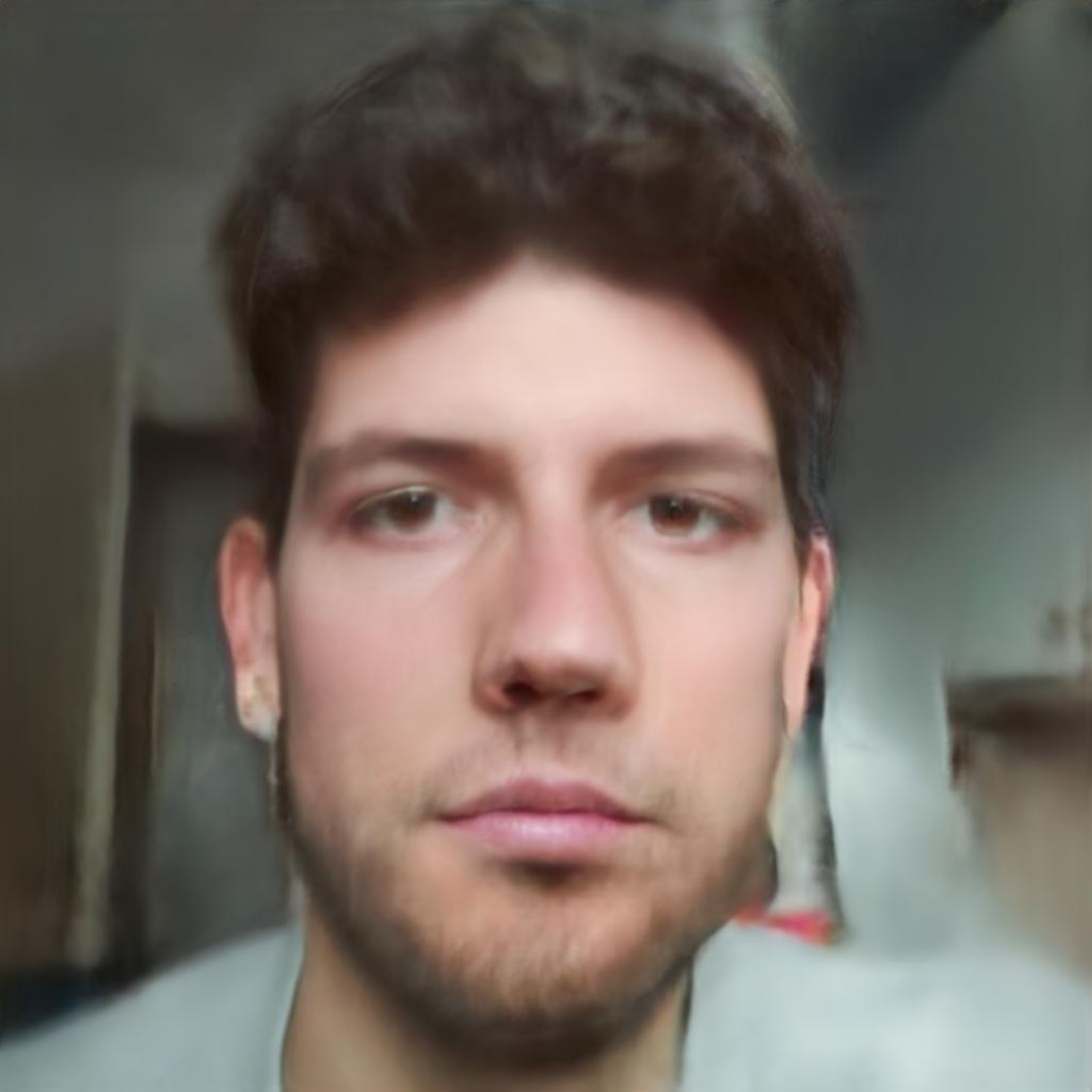}
\end{subfigure}
\begin{subfigure}[b]{0.16\textwidth}
\includegraphics[width=\textwidth]{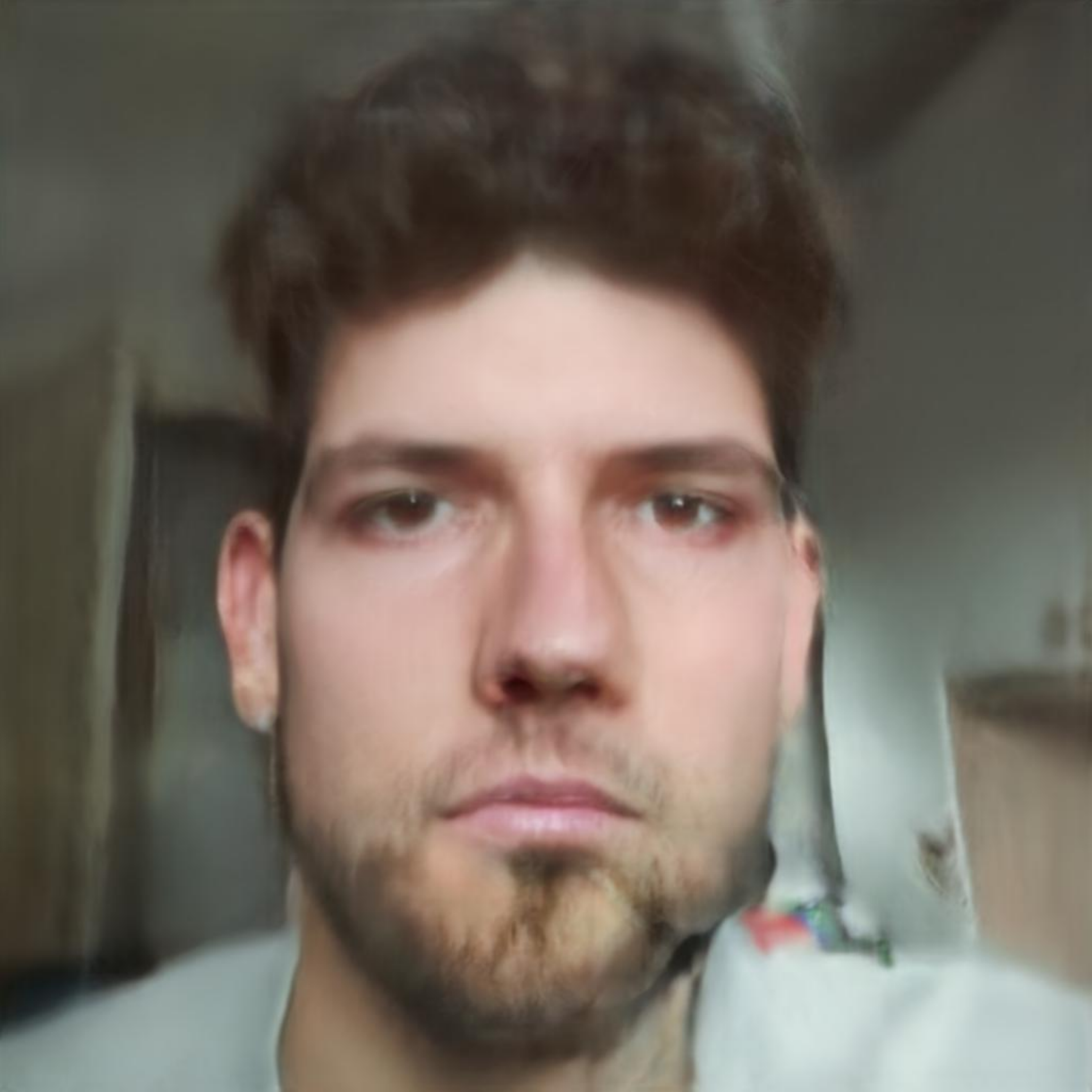}
\end{subfigure}
\begin{subfigure}[b]{0.16\textwidth}
\includegraphics[width=\textwidth]{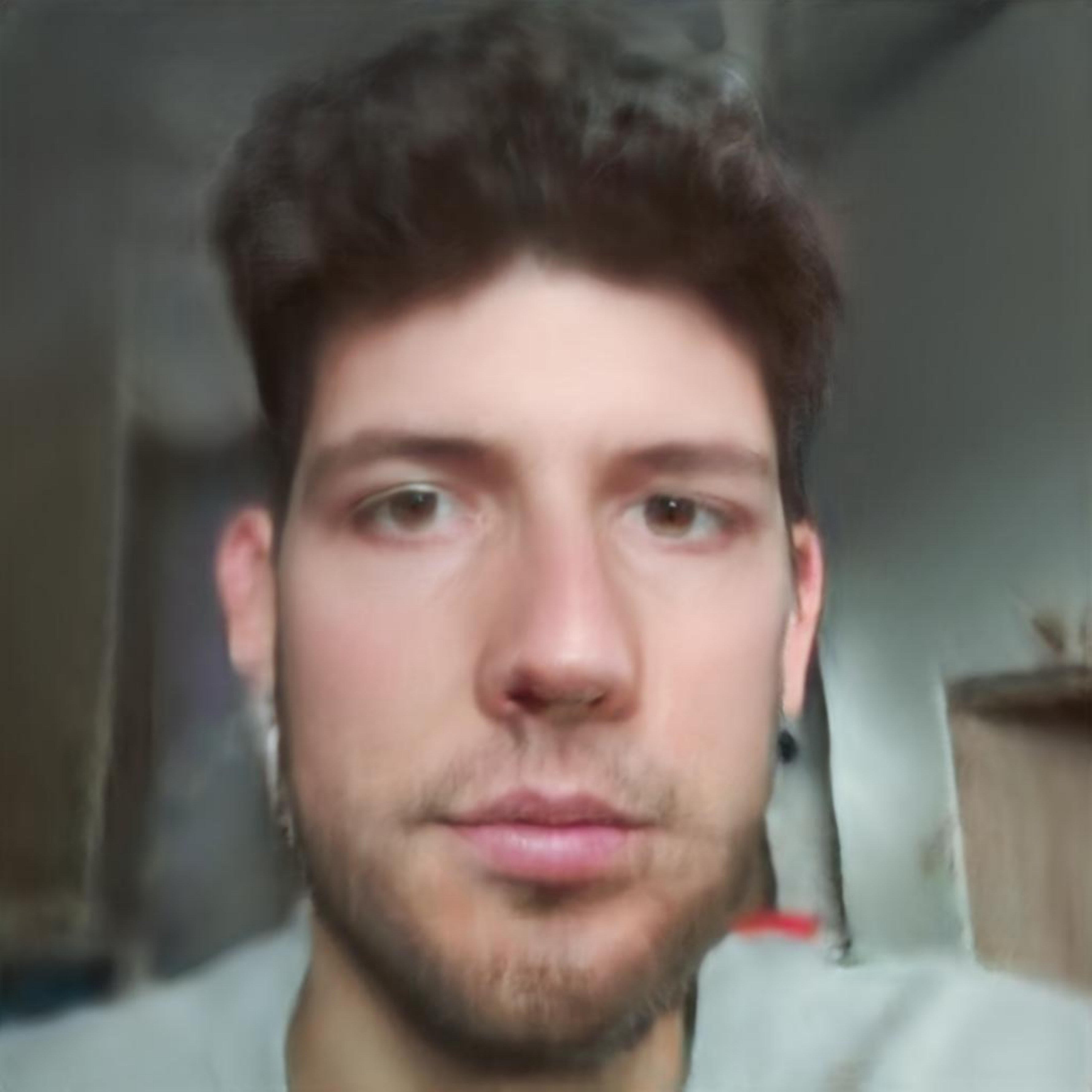}
\end{subfigure}

\caption{Top row: solutions found by PULSE ($\ell_2$ range: $[1.3\times10^{-4},1.5\times10^{-4}]$). Mid row: solutions found by using $\v^8$ and $\v^{26}$ as directions ($\ell_2$ range: $[3.9\times10^{-3},4.7\times10^{-3}]$). Bottom row: solutions found by optimized $\d$ as direction ($\ell_2$ range: $[1.2\times10^{-3},2\times10^{-3}]$).}
\label{fig:IP_solutions}
\end{figure*}

Fig. \ref{fig:IP_solutions} shows the results for the IP problem. 
Even in this case, PULSE (top row) found solutions with different perceptual changes, but still some of them seem to introduce unnatural variations, like the one on the nose or some distortion on the mouth. The second row shows images along two starting eigenvectors of the Hessians without applying our method. Although notable perceptual changes are visible, the $\ell_2$ distance on the measure is outside of our constraint and indeed some differences on the eyes are created with respect to the observations. Finally, the last row, exploits our optimized direction showing how the $\ell_2$ distance on the measurements is decreased while retaining good semantic changes in the masked area, such variations in the lip thickness or the amount of beard.

We remark that PULSE may be able to find good solutions but it has two main drawbacks that are solved by the proposed technique. First, it lacks any explicit control on the measurements distance. Constraining the $\ell_2$ distance between the original measurements and the measurements of the reconstruction to a feasibility threshold can only be done by enforcing a stopping criterion on the inversion optimization problem. However, due to the non-convex nature of the problem this often results in degenerate solutions that no longer belong to the manifold of realistic images like some of the ones previously shown.  

Another advantage of the proposed method with respect to PULSE is the computational complexity due to PULSE requiring to solve a full optimization problem to generate a new solution. For our proposed method, this needs to be only done once, coupled with the estimation of the Hessians, but then multiple solutions can be generated almost instantaneously. Table \ref{table:time} reports the time required for the two methods to generate ten solutions. This time does not account for bad solutions: indeed, discarding bad minima found by PULSE would further increase its computational requirements.

\begin{table}
\setlength{\tabcolsep}{3pt}
\centering
 \caption{Computational time required to find 10 solutions, with our method and using different initializations (PULSE).}
 \vspace*{-6pt}
 \begin{tabular}{lccc}
    & Inverse Problem & Model & Time (s) to 10 solutions \\
  \hline
  \hline
  PULSE    &  SR  & BigGAN & $3.5 \times 10^{4}$\\
  \hline
  \textbf{e-GLASS}    &  SR  & BigGAN & $\boldsymbol{ 3.7 \times 10^{3}}$\\
  \hline
  \hline
  PULSE    &  IP  & PGGAN & $7.9 \times 10^{3}$\\
  \hline
  \textbf{e-GLASS}    &  IP  & PGGAN & $\boldsymbol{1.3 \times 10^{3}}$\\
  \hline
 \end{tabular}
 \label{table:time}
 \vspace*{-9pt}
\end{table}

\vspace*{-5pt}
\section{Conclusions}
\label{sec:conclusions}
\vspace*{-5pt}

We proposed a novel framework to explore the solution space of linear inverse problems. By exploiting the power of the geometry of the latent space around an initial solution, we showed it is possible to search for novel solutions that have semantic differences between each other. We confirmed 
the quality and variety of the multiple solutions as well as the reduced complexity when compared with state-of-the-art iterative optimization methods on the SR and IP problems.

\clearpage

\ninept
\bibliographystyle{IEEEbib}
\bibliography{biblio}

\end{document}